\documentclass[fleqn,usenatbib]{mnras}

\usepackage{newtxtext,newtxmath}

\usepackage[T1]{fontenc}
\usepackage{ae,aecompl}

\usepackage{graphicx}	
\usepackage{amsmath}	
\usepackage{amssymb}	
\usepackage{threeparttable}

\newcommand{\msuneq}{M_{\odot}}
\newcommand{\rvir}{$r_{vir}$}

\newcommand{\cfour}{C~\textsc{iv}}
\newcommand{\lya}{Ly$\alpha$}
\newcommand{\lyb}{Ly$\beta$}
\newcommand{\osix}{O~\textsc{vi}}

\newcommand{\cthree}{C~\textsc{iii}}

\newcommand{\sithree}{Si~\textsc{iii}}
 
\newcommand{\hone}{H~\textsc{i}} 
\newcommand{\cmt}{{\rm cm}^{-2}}

\title[UV Signature of Multiphase ICM]
{Ultraviolet Signatures of the Multiphase Intracluster and Circumgalactic Media in the RomulusC Simulation}

\author[Butsky et al.]{
Iryna S. Butsky,$^{1}$\thanks{E-mail: ibutsky@uw.edu}
Joseph N. Burchett,$^{2}$
Daisuke Nagai,$^{3-5}$
Michael Tremmel,$^{3,5}$
\newauthor 
Thomas R. Quinn,$^{1}$
and
Jessica K. Werk $^{1}$
\\
$^{1}$ Department of Astronomy, University of Washington, Seattle, WA, 98195-1580, USA\\
$^{2}$ Department of Astronomy and Astrophysics, University of California, Santa Cruz, USA\\
$^{3}$ Department of Physics, Yale University, New Haven, CT 06520, USA\\
$^{4}$ Department of Astronomy, Yale University, New Haven, CT 06520, USA\\
$^{5}$ Yale Center for Astronomy and Astrophysics, New Haven, CT 06520, USA
}

\date{Accepted XXX. Received YYY; in original form ZZZ}

\pubyear{2015}

\begin{document}
\label{firstpage}
\pagerange{\pageref{firstpage}--\pageref{lastpage}}
\maketitle

\begin{abstract}

Quasar absorption-line studies in the ultraviolet (UV) can uniquely probe the nature of the multiphase cool-warm ($10^4<\mathrm{T}<10^6$K) gas in and around galaxy clusters, promising to provide unprecedented insights into 1) interactions between the  circumgalactic medium (CGM) associated with infalling galaxies and the hot ($\mathrm{T}>10^6$K) X-ray emitting intracluster medium (ICM), 2) the stripping of metal-rich gas from the CGM, and 3) a multiphase structure of the ICM with a wide range of temperatures and metallicities. In this work, we present results from a high-resolution simulation of a $\sim 10^{14} \msuneq$ galaxy cluster to study the physical properties and observable signatures of this cool-warm gas in galaxy clusters.  We show that the ICM becomes increasingly multiphased at large radii, with the cool-warm gas becoming dominant in cluster outskirts. The diffuse cool-warm gas also exhibits a wider range of metallicity than the hot X-ray emitting gas. We make predictions for the covering fractions of key absorption-line tracers, both in the ICM and in the CGM of cluster galaxies, typically observed with the Cosmic Origins Spectrograph aboard the Hubble Space Telescope (HST).  We further extract synthetic spectra to demonstrate the feasibility of detecting and characterizing the thermal, kinematic, and chemical composition of the cool-warm gas using \hone, \osix, and \cfour\ lines, and we predict an enhanced population of broad \lya\ absorbers tracing the warm gas. Lastly, we discuss future prospects of probing the multiphase structure of the ICM beyond HST.

\end{abstract}

\begin{keywords}
galaxies:clusters:general -- galaxies:clusters:intracluster medium -- galaxies:halos -- methods:numerical
\end{keywords}



\section{Introduction}
Galaxy clusters form ecosystems on the largest mass and spatial scales and serve as unparalleled laboratories for galaxy evolution and extreme magnetohydrodynamic phenomena. Clusters are marked by hot, X-ray emitting, intracluster media (ICM) that serve as records of cluster dynamics and chemical enrichment history.  All infalling galaxies are subject to the punishing conditions of the ICM throughout their journey, while individual galaxies are embedded in diffuse, multiphase circumgalactic media (CGM) that mediate the `baryon cycle', consisting of galactic outflows and accretions of gas that ultimately dictate star formation activity and the transport of heavy elements \citep{tumlinson17}.  Thus, it is the interplay between the ICM and CGM that manifests in many aspects of cluster gas physics, and understanding the complex thermodynamic and chemical structures of the ICM is critical to understanding the evolution of cluster galaxies \citep[see][for reviews]{Kravtsov:2012, Walker:2019}.

Galaxy clusters are also sites of extraordinary galaxy transformation. On average, cluster galaxies are more likely to be quenched and to have less gas in their interstellar medium (ISM) \citep{Davies:1973, Dressler:1980, Sandage:1985} than isolated field galaxies.  There are several ways in which the cluster environment may contribute to this gas shortage, although the dominant mechanisms are still debated \citep{Fraser:2017}: a combination of tidal stripping by the massive center galaxy \citep{Moore:1996, Lake:1998, Wang:2004}, high-speed fly-bys, and mergers with other galaxies \citep{Toomre:1972, White:1978, Roediger:2015} or ram-pressure stripping from the ICM \citep{Gunn:1972, Fumagalli:2014}. This cool, metal-rich, stripped gas contributes to the clumpy morphology in cluster outskirts before ultimately mixing with the ICM \citep{Tonnesen:2007, Jachym:2014}.  Recent studies have also shown that the CGM of cluster galaxies is quite deficient relative their more isolated counterparts \citep{Yoon:2013,Burchett:2018aa}, although the samples remain small and it is less clear where this transformation occurs.  If galaxies' CGM are stripped early upon infall, this may lead to a much slower quenching process than rapid stripping of the interstellar medium (ISM) \citep{Zinger:2018aa}.

To date, X-ray and microwave measurements of the thermodynamic and metallicity profiles of the ICM have been critical towards determining the physics governing structure formation \citep[][for reviews]{Walker:2019, Mroczkowski:2019aa}, chemical enrichment of the ICM \citep[][for a review]{Werner:2013,Urban:2017,Mernier:2018ab}, and cosmological models \citep[][for reviews]{Allen:2011aa,Pratt:2019aa}. Modern cosmological simulations of galaxy cluster formation predict that the outskirts of galaxy clusters is a cosmic-melting pot, where the structure formation processes, such as mergers and gas accretion \citep{Lau2015,Zinger2016}, generate clumpy \citep{Nagai:2011,Zhuravleva2013,Vazza:2013aa,Rasia2014} and turbulent \citep{Lau:2009qy,Nagai:2013aa,Nelson:2014aa} hot X-ray emitting ICM in the outskirts of galaxy clusters. In this work, we explore an alternative approach to studying the ICM and cluster galaxy CGM: probing the diffuse, cool-warm, gas that is virtually invisible to the observational techniques described above, using using absorption-line (UV) spectroscopy.

Rest-frame UV absorption-line spectroscopy studies have revolutionized our understanding of how the CGM influences and, conversely, is influenced by relatively isolated galaxies \citep[e.g.][]{Tumlinson:2011, tripp11, Stocke:2013, Werk:2013, Bordoloi:2014}. Our picture of the CGM has transformed from that of a quiescent remnant of galaxy formation, to one of a turbulent, multiphase medium with a rich ionization and kinematic structure that plays a key role in galaxy evolution \citep{tumlinson17}. Similarly, UV observations of the cool-warm ICM have the potential to provide unprecedented insights to the thermodynamic, kinematic, and chemical properties of galaxy clusters. Directly observing the stripping of the CGM of infalling galaxies will place stringent constraints on 1) the dominant quenching mechanisms in high density environments, 2) the conditions and timescales that govern CGM stripping through interactions with the ICM, 3) the distribution of metals across all phases of the ICM, and 4) the role of feedback on the thermodynamic and chemical properties of the CGM. Additionally, UV-derived column density measurements and photoionization modeling can constrain the baryon fraction in the hot ($\mathrm{T} > 10^6$K) X-ray emitting ICM to complement those estimates made using X-ray and microwave observations \citep{Burchett:2018aa,Ge:2016lr,Wang:2014lr}. 

Several groups have successfully pioneered UV observations of the cool-warm gas in galaxy groups and clusters using \emph{HST/COS} \citep[e.g.,][]{Yoon:2012yu, Johnson:2015, Muzahid:2017lr, Pointon:2017, Yoon:2017aa, Burchett:2018aa, Nielsen:2018, Stocke:2019}, although the large samples of UV sightlines systematically probing cluster environments have not yet been obtained as exist for CGM samples. Next-generation UV telescopes \citep[e.g., \emph{LUVOIR};][]{Bolcar:2017aa} have the potential to increase both the number of observable background sources (e.g., quasars (QSOs) and galaxies) and the sensitivity to lower column densities of cool-warm gas by orders of magnitude. We return to compare current and future capabilities later in Section \ref{sec:prospects}.

In this work, we characterize the multiphase ICM using a high resolution simulation of a $\sim 10^{14}\msuneq$ cluster and make testable predictions for its observable properties, both with \emph{HST-COS} and with future instruments. This work is organized as follows. In \S\ref{sec:methods}, we describe our methods, the {\sc RomulusC} simulation of a galaxy cluster, and the tools integral to our analysis.  In \S\ref{sec:multiphaseICM}, we present evidence for a multiphase, metal-enriched ICM. In \S\ref{sec:CGMinClusterGals}, we discuss the properties of the CGM of cluster galaxies. In \S\ref{sec:specAnalysis}, we demonstrate how observable signatures trace the physical conditions of the gas using synthetic spectra. In \S\ref{sec:prospects}, we discuss the current and future prospects of observing the cool-warm gas in galaxy clusters. We summarize our results in \S\ref{sec:conclusions}.

Throughout this work, we divide the gas into four distinct phases: {\it cold} ($\mathrm{T}<10^4$K), {\it cool} ($10^4<\mathrm{T}<10^5$K), {\it warm} ($10^5<\mathrm{T}<10^6$K) and {\it hot} ($\mathrm{T} >10^6$K).  This convention differs somewhat from that adopted elsewhere in the literature \citep[e.g.,][]{Cen:1999yq}, as we simply refer to the $10^5<\mathrm{T}<10^6$K phase as `warm' rather than `warm-hot'.  In the context of galaxy clusters, which contain an abundance of $\mathrm{T}>10^6$K gas that is often simply deemed `hot' in the literature, we feel that our nomenclature avoids ambiguity when discussing relationships between the phases.  Thus, in this paper, we focus primarily on gas at cool-warm temperatures.

\begin{figure*}
	\includegraphics[width=\textwidth]{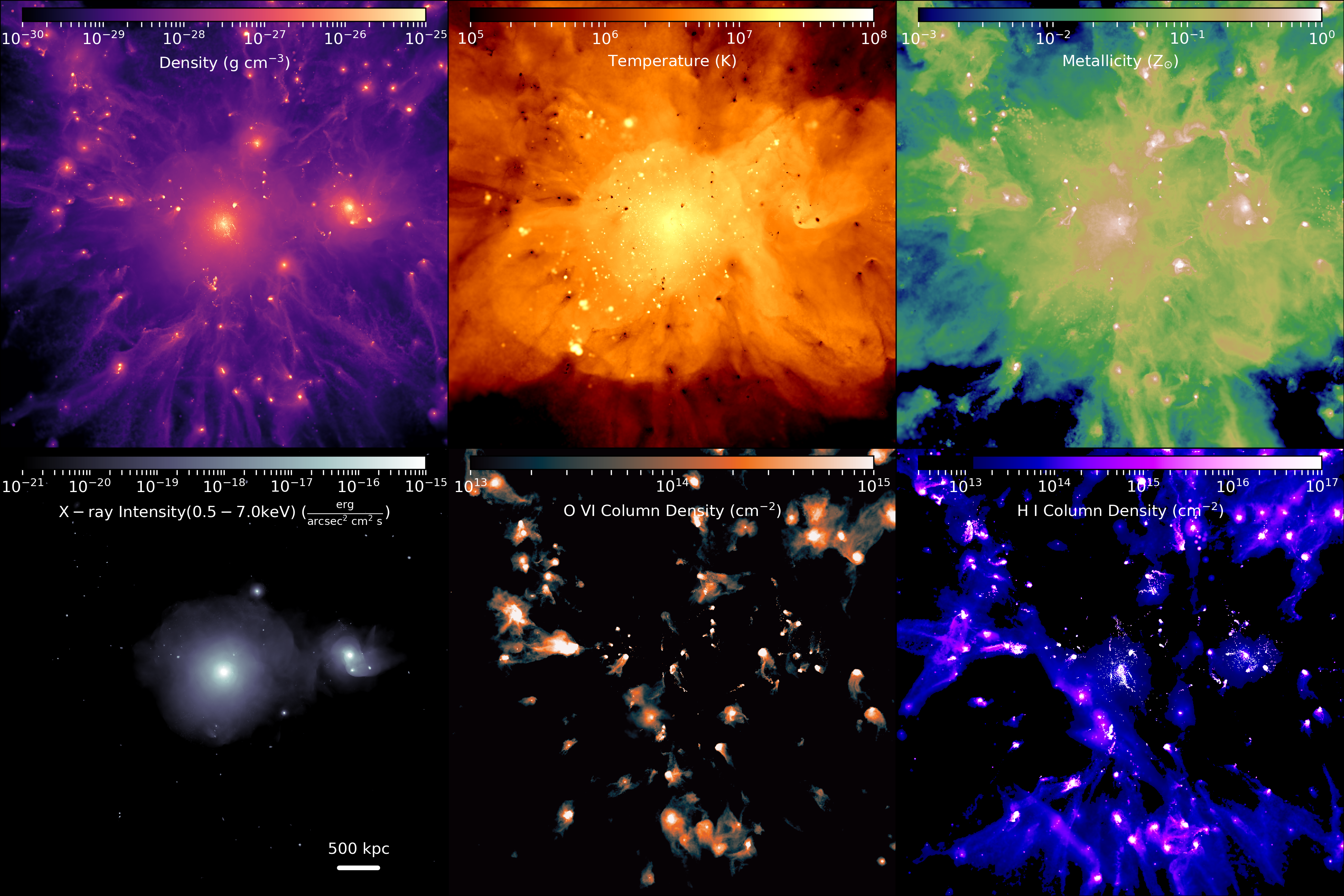}

    \caption{A 5 x 5 Mpc snapshot (in physical units) of the {\sc RomulusC} simulation at
    $z = 0.31$. {\it Top row:} The density-weighted projections of gas density, temperature, and metallicity. 
    {\it Bottom row:} The integrated X-ray intensity, \osix\ column density, and \hone\ column density.
   The color scaling of the images in the bottom row is roughly consistent with the sensitivity of existing instruments.  
   Compared to X-ray emission, the structure probed by UV absorption studies (\osix\ and \hone\ column densities) better traces cluster
   outskirts and small-scale structure within the virial radius.$^1$}

	\label{fig:multipanel}
\end{figure*}

\section{Methods}
\label{sec:methods}

\subsection{The RomulusC Simulation}

{\sc RomulusC} \citep{Tremmel:2019} is a hydrodynamic cosmological zoom-in simulation of a $10^{14}$ M$_{\odot}$ galaxy cluster run with the new Tree+Smoothed Particle Hydrodynamics (SPH) code, {\sc ChaNGa}  \citep{Menon:2015}. The zoom-in region is a Lagrangian region derived from the highest mass halo selected from a 50 Mpc uniform volume, dark matter-only simulation.  {\sc RomulusC} is among the highest resolution hydrodynamic cosmological simulations of a galaxy cluster run to date, with a Plummer equivalent gravitational force resolution of 250 pc (a 350 pc spline kernel is used), a maximum SPH resolution of 70 pc, and gas and dark matter mass resolutions of $2.12\times10^5$ and $3.39\times10^5M_{\odot}$ respectively. At this resolution, {\sc RomulusC} is one of only two cluster cosmological simulations currently in production that is capable of resolving the multiphase structure of gas in the ICM (see also \cite{Nelson:2019}). Lower mass halos and galaxies, including their interaction with the ICM, are also followed with greater detail than ever before. {\sc RomulusC} resolves halos as small as $3\times10^9$ M$_{\odot}$ with at least $10^4$ particles (spatial resolution sufficient for resolving ram-pressure stripping \citep{Roediger:2015}). As the CGM and ISM of infalling galaxies also contributes to the temperature, density, and metallicity distribution of gas within the ICM, {\sc RomulusC} is able to provide a more complete picture of the ICM structure than other cluster simulations with more typical resolutions (often lower by factors of 10 or more in both mass and spatial resolution, see Table 1 in \citet{Tremmel:2019}).

{\sc ChaNGa} uses sub-grid physics modules for star formation, stellar feedback, UV background with self-shielding, and low temperature metal cooling, that have been previously implemented and tested in {\sc Gasoline} \citep{Wadsley:2004, Wadsley:2008, Wadsley:2017}. Supermassive black hole (SMBH) formation, dynamics, growth, and feedback is also included \citep{Tremmel:2015, Tremmel:2017}. Importantly, the simulation includes an updated implementation of turbulent diffusion \citep{Wadsley:2017}, shown to be important for producing realistic entropy profiles in galaxy cluster cores \citep{Wadsley:2008} and metal distributions in galaxies \citep{Shen:2010}. The combination of a gradient-based shock detector, time-dependent artificial viscosity, and an on-the-fly time-step adjustment system allows for a more realistic treatment of both weak and strong shocks \citep{Wadsley:2017}. Both the thermal and turbulent diffusion of metals is accounted for following \citet{Shen:2010}.
\footnotetext[1]{While a number of `virial radius' definitions exist in the literature, we generally adopt \rvir~=~$R_{200c}$, the radius within which the average density is 200x the critical density of the Universe.  Other definitions may appear in the literature, and we provide the following the approximate scaling relation:  $R_{500c}$ : $R_{200c}$ : $R_{200m}$ = 1 : 1.4 : 3.} 
Feedback from both stars and SMBHs is followed throughout the simulation. As in previous work for runs at this resolution \citep{Stinson:2006}, star formation is assumed to occur with an efficiency of 15\% in dense ($>0.2$ cm$^{-3}$), cold ($\mathrm{T}<10^4$K) gas on a characteristic timescale of $10^6$ years. Supernova (SN) feedback follows the ``blastwave'' approach \citep{Stinson:2006}, with each SN thermally coupling $0.75 \times 10^{51}$ erg to the surrounding gas. 

SMBHs are seeded based on local gas properties, assumed to form in very high density ($n_{gas} >3 \,\mathrm{m}_{\mathrm{H}}/ \mathrm{cm}^{3}$), low metallicity ($Z<3\times10^{-4}Z_{\odot})$ regions, which are self-consistently predicted to occur in the centers of low mass ($10^8-10^{10}$ M$_{\odot}$) halos in the early Universe ($z>5$). SMBHs grow through both mergers and accretion of gas via a modified Bondi-Hoyle prescription that accounts for both additional support from angular momentum  \citep{Tremmel:2017} and the unresolved multiphase structure in the ISM \citep{Booth:2009}. A small fraction of the mass-energy of accreted material (0.2\%) is transferred to nearby gas via thermal feedback and naturally results in large-scale outflows and has been shown to successfully regulate, and sometimes fully quench, star formation in massive galaxies \citep{Pontzen:2017, Tremmel:2017, Tremmel:2019}. 

The sub-grid models used here were derived from an extensive parameter calibration program, ensuring that the {\sc Romulus} simulations create galaxies with realistic stellar masses, SMBH masses, gas content, and angular momentum \citep[see][for more details]{Tremmel:2017}. 
Importantly, the sub-grid physics used in the {\sc Romulus} simulations is tuned to reproduce realistic galaxy properties up to Milky Way-mass halos. Despite this, the simulation successfully predicts realistic SMBH and stellar masses for halos up to $2\times10^{13}$ M$_{\odot}$. No effort was made to optimize the simulation to produce a realistic galaxy cluster, so all of the results from {\sc RomulusC} are predictions from the simulation.

Note that {\sc RomulusC} does not include metal-line cooling for temperatures above $10^4$K. This is one of the limitations of this simulation, as metal lines are an important coolant for warm-hot gas in the ICM. More specifically, at the gas temperatures ($\mathrm{T}\sim10^5$K) and metallicities ($\sim0.1-0.3 Z_{\odot}$) most relevant in this work, the extragalactic UV background dominates the heating/cooling balance, so the lack of metal-line cooling will result in cooling times that are a factor of $\sim3-5$ too long \citep{Shen:2010}. Furthermore, cooling in 10$^{4-5}$K gas is dominated by the He Ly$\alpha$ line, and primordial cooling does a reasonably good job of estimating metal-line cooling \citep{Hopkins:2018}. We do include metal-line cooling for gas below $10^4$ K, where contribution from metals becomes important. 

At the cluster outskirts, where much of this work is focused, the cooling times are already quite long, even if metal cooling was accounted for, hence the inclusion of metal line cooling does not affect accretion of gas onto massive halos \citep{vandeVoort:2011}. The content of ions such as \osix\ in the ICM of {\sc RomulusC} should therefore be robust to well within an order of magnitude. Closer to the BCG (and infalling cluster galaxies), the gas is just as sensitive to feedback physics as to the cooling model used \citep{vandeVoort:2011}. It is in this regime where the halo gas is strongly coupled to the ISM via feedback processes and it is not possible to separate the effects of high temperature cooling from feedback (and, therefore, ISM physics). Therefore, the robustness of the gas properties in the simulation is difficult to estimate.

At $z=0$, the cluster is experiencing an on-going head-on merger with 1:8 mass ratio. The infalling halo is still outside $R_{200}$ at $z=0.3$ and begins to have an effect on the state of the ICM in the inner cluster at $z \lesssim 0.2$. The merger results in the destruction of the cool-core of the cluster (Chadayammuri et al., in prep). For most of the following analysis, in order to avoid the effects of a transient event, we focus on the state of the ICM when the cluster is still relaxed at $z=0.31$, though we note that this is a higher redshift than the available observations we use for comparison. The virial radius of the cluster at this redshift is 810 kpc. The zoom-in Lagrangian region of {\sc RomulusC} extends to $\sim3$ Mpc from cluster center at $z=0.3$. 
Since the physical state of the gas near the outskirts of the Lagrangian region are subject to numerical effects, we limit our analysis to within this region. Hence, we do not expect this to affect any of our conclusions.

\begin{figure}
	\includegraphics[width=\columnwidth]{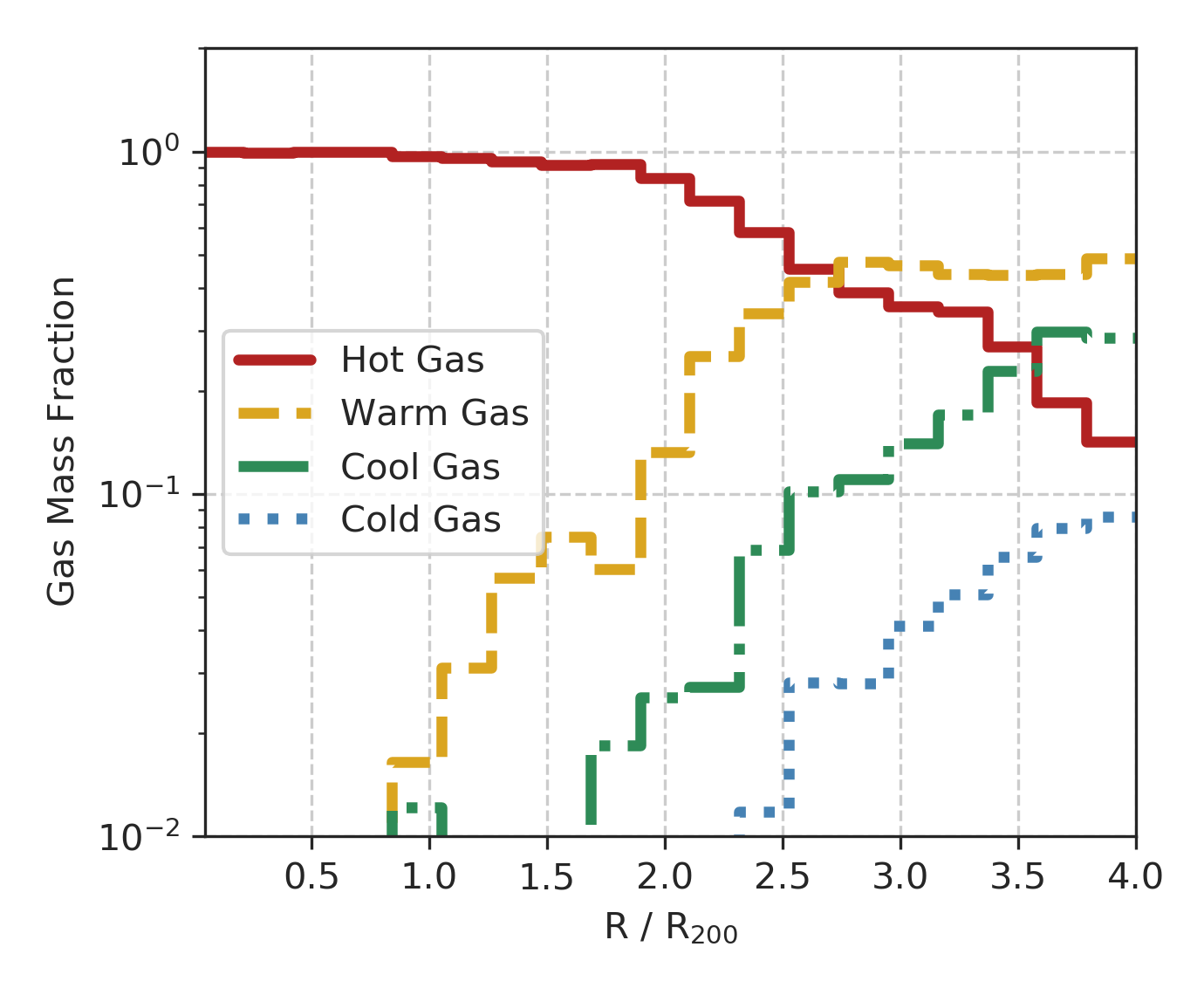}
    \caption{The mass fraction of cluster gas in the cold ($\mathrm{T}< 10^4$K), cool ($10^4 < \mathrm{T} < 10^5$K), warm ($10^5 < \mathrm{T} < 10^6$K),and hot ($\mathrm{T} > 10^6$K) phases as a function of 3D distance from the cluster
    center in the {\sc RomulusC} simulation at $z=0.31$. Although hot gas dominates within the virial radius of the cluster, 
    warm and cool gas comprise the majority of the gas mass at the cluster outskirts. }
    \label{fig:gas_fraction} 
\end{figure}
\begin{figure*}
\begin{center}
	\includegraphics[width=\textwidth]{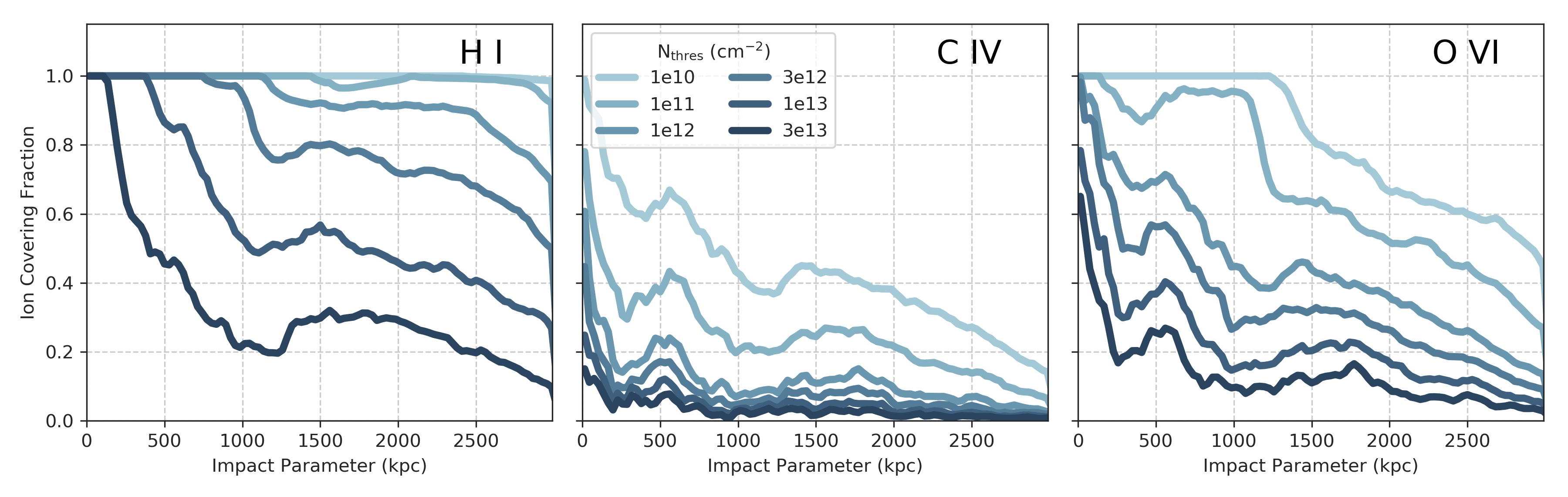}

    \caption{ The covering fraction of \hone\ ({\it left panel}), \cfour\ ({\it middle panel}), and \osix\ ({\it right panel}) 
    as a function of the impact parameter measured from the cluster center at $z = 0.31$. Each line represents 
    the covering fraction for different observational column density thresholds. With existing observational thresholds of
    \emph{HST/COS} 
    ($10^{13} \rm {cm}^{-2}$ for \hone, $10^{13.1} \mathrm{cm}^{-2}$ for \cfour\, and  $10^{13.3} \mathrm{cm}^{-2}$ for \osix\ ),
    we can expect covering fractions above 0.5 and 0.15 for \hone\ and \osix\ respectively within 1.5 Mpc of the cluster center.
    At existing observational column density sensitivities, the expected covering fraction for \cfour, is below 0.1. 
    Improving the observational threshold by an order of magnitude would result in roughly a 2x increase in the covering 
    fraction of \hone\ and \osix. 
 }    
    \label{fig:cfrac_threshold}
\end{center}
\end{figure*}

\subsection{Ion Fields and Synthetic Absorption Spectra}
\label{sec:synthSpec}

We analyze our simulation data with a combination of {\sc yt} \citep{Turk:2011}, {\sc Trident} \citep{Hummels:2017}, 
{\sc pynbody} \citep{pynbody}, and {\sc Tangos} \citep{Pontzen:2018}. In this section, we describe the process of transforming the physical quantities in simulation outputs to observable quantities and, ultimately, synthetic spectra. 

Keeping track of individual atomic elements throughout the simulation is computationally expensive.
Although some zoom-in simulations of low-mass galaxies have successfully tracked over 11 individual elements \citep[e.g.,][]{Oppenheimer:2013, Emerick:2019}, such
detailed tracking is computationally infeasible, with current technologies, at the cluster scale. 
Instead, {\sc RomulusC} explicitly tracks hydrogen, oxygen, and iron.
The metallicity used here is a weighted sum of oxygen and iron based on solar abundance ratios \citep{Asplund:2009}.

From the metallicity, we calculate specific ion abundances and their ionization states using
{\sc Trident}\footnote{Trident is built as an extension of the widely-used analysis tool, {\sc yt} \citep{Turk:2011} and is compatible with most astrophysical simulation codes.},
the first open-source, standardized tool for determining ion abundances and generating synthetic spectra from simulations.
{\sc Trident} derives the abundance of each element in the simulation, $n_X$, by using the local metallicity and assuming a solar abundance by number
\footnote{$n_X = n_{\mathrm{H}} Z (n_x/n_{\mathrm{H}})_{\odot}$, where $n_{\mathrm{H}}$ and $n_X$ are the number densities of Hydrogen and any element, X, $Z$ is the metallicity, 
and the subscript $\odot$ indicates the solar ratio}. 
Then, the collisional ionization of each element is determined from the local gas temperature, density, and metallicity,
using an extensive pre-tabulated look-up table generated by {\sc CLOUDY} \citep{Ferland:2013}. The photionization is calculated assuming an extragalactic UV background
described in \citet{HaardtMadau:2012}. Using this method, we can determine the mass and number density of any ion throughout the simulation. 

We note that we do not explicitly include photoionization from local sources (such as nearby galaxies or AGN). 
This could produce a significant variation in the photionization and can have a large impact on ion 
abundances \citep{Oppenheimer:2018}. The uncertainty in the assumed UV background is the largest source of 
uncertainty in the relative ionic abundances that we derive. For a detailed account of the impact of a varying UVB slope on ion abundances, see the appendix of \citet{Werk:2014}.

To compute column density profiles, we first generate column density projections centered on the cluster 
(or individual cluster galaxy) along each axis. From these projections, we gather the distribution of column densities as a function
of impact parameter. To convert these column density profiles to covering fractions, we count the fraction of ``sightlines'' in each
radial bin that is above some threshold column density. The existing column density thresholds of HST-COS are $10^{13}, 10^{13.1}, 10^{13.3}$ $\rm{cm}^{-2}$ for \hone, \cfour, and \osix\ respectively.

We also use {\sc Trident} to generate synthetic spectra, allowing for direct comparison with existing observations. 
To generate a spectrum, we first designate an arbitrary sight-line through the simulation volume, analogous to probing
a foreground galaxy along the line-of-sight to a QSO. The sightline, $l$, connects points a and b in a simulation such that 
$\vec{l} = \vec{r_b} - \vec{r_a} \sum^n_{i=0} d\vec{l_i}$. The length of vector element, $d\vec{l}$ is set by the resolution of
the simulation. Absorption features are calculated based on the species' abundance and gas properties at each vector element. 
To make the synthetic spectra more similar to realistic spectra, {\sc Trident} adds instrument-specific line-spread functions, 
a Milky Way foreground, a background QSO spectrum, and variable signal-to-noise ratio (S/N).   For the analysis presented in \S\ref{sec:specAnalysis}, we extract synthetic spectra using the \emph{HST/COS} line-spread function \citep{COSLSF}, 
wavelength coverage of the COS G130M and G160M gratings (approximately 1150-1800\AA), no Milky Way foreground absorption, and S/N = 10.  
These choices are motivated by the feasibility with which our theoretical
predictions may be observationally constrained (see discussion in \S\ref{sec:prospects}).

\section{Multiphase ICM Properties in the RomulusC simulation}
\label{sec:multiphaseICM}

\subsection{Evidence for the multiphase ICM} 

We first demonstrate that the ICM in our simulation is, in fact, multiphase: that gas clouds spanning orders of magnitude in temperature and density exist co-spatially. 
The high resolution of {\sc RomulusC} is well-suited for resolving the multiphase ICM. With higher resolution (more particles
per volume) each spatial region in the simulation is represented with a wider distribution of temperatures and densities, 
so that the gas is more accurately divided into varying thermodynamic phases \citep{Hummels:2018}.

Figure \ref{fig:multipanel} captures the state of the {\sc RomulusC} cluster at $z = 0.31$. 
The top row highlights properties that are directly measured from simulations but can only be inferred from observations: the density, temperature, and metallicity. 
The density projection highlights the physical structure of the cluster filled with numerous satellite galaxies. 
The most massive halo is at the center of the projection frame, surrounded by the hottest and densest regions of the ICM. 
This cluster is about to undergo a major merger with another galaxy group, which is approaching from the right. Dense filaments trace galaxies flowing through the medium as well as gas being stripped from these galaxies through ram-pressure stripping. 

The temperature profile is dominated by hot gas at the center of the cluster, reaching ambient temperatures above $10^7$K.  Although rare near the cluster center, filaments of warm gas are visible just outside of the virial radius (also see Figure \ref{fig:gas_fraction}). These gaseous filaments are likely a combination of heated cool gas inflows and stripped material from infalling galaxies. 

The projected metallicity in the top-right panel demonstrates the extent to which the ICM is enriched with heavy elements. The ambient ICM near
the cluster center has a density-weighted average metallicity around $0.3Z_{\odot}$, roughly consistent with values inferred from X-ray observations \citep{Urban:2017}. 
However, the metallicity distribution is far from uniform. At the cluster outskirts, the projected metallicity spans several orders of magnitude (see Figures \ref{fig:metallicity} and \ref{fig:metallicity_radius} for more quantitative analyses of the distribution of metals). The high-metallicity regions trace infalling cluster galaxies, and these
structures are easily observable in \hone\ and \osix\ (see bottom row). The low-metallicity regions show the more pristine ICM, which is difficult to observe with both X-ray 
emission and UV absorption. The metallicity distribution gradually becomes less clumpy towards the center of the ICM, 
where gas has had more time to mix. Understanding this distribution of metals throughout the cluster constrains the motion of galaxies through the ICM and the mechanism through which they lose gas.

While the density, temperature, and metallicity are readily available from the simulation data, these properties cannot be measured directly using observations. Rather, these properties must be inferred through careful modeling, which itself depends on various assumptions and model parameters. For this reason, we also show the ``observable" cluster properties in the bottom row of Figure \ref{fig:multipanel}. The left panel shows 
the expected X-ray intensity in the 0.5-7.0 keV band and the subsequent middle and right panels show the \osix\ and \hone\ column densities, respectively. 
The limits on the colorbars are chosen such that the black background roughly indicates the observational limits of existing instruments, such as \emph{Chandra}, \emph{XMM-Newton} and \emph{HST/COS} \citep[for sensitivities typically obtained, e.g., COS-Halos;][]{Tumlinson:2011}. 

The gaseous structure observed with X-rays traces the hot, dense gas of the ICM. In contrast, the regions traced through UV absorption
comprise the cool-warm phases of the CGM.  Warm gas, traced by the \osix\ column density (bottom row, middle panel), exists as a combination of a volume-filling halo gas and denser gas associated with galaxies. The volume-filling warm gas produces \osix\ column densities around $10^{13} \mathrm{cm}^{-2}$ in halos of cluster galaxies, especially at the cluster outskirts. Near the centers of gas-rich cluster galaxies, the \osix\ column density consistently exceeds $10^{16} \mathrm{cm}^{-2}$. Galaxies moving through the ICM at sufficiently high velocities are stripped of their gas due to ram-pressure forces. Several galaxies in Figure \ref{fig:multipanel} are oriented so that their gas-stripped tails are bright in \osix.  

The cool gas, traced by the column density of \hone\ (bottom panel), paints a similar picture for the multiphase structure in the ICM. Compared to the warm gas, the cool gas is more concentrated in galactic centers, and the stripped gas tails of infalling galaxies appear narrower. In certain cases, the \hone\ shows small-scale, dense structure that is washed out in the distribution of \osix. The diffuse cool gas survives in abundance, both near the cluster center and well beyond its virial radius. The overall similarity in the structures traced by both the warm and cool gas demonstrates the existence of a multiphase medium. 

\subsection{Radial distribution of the multiphase ICM}
We now quantify the mass distribution of the multiphase gas in the cluster at $z = 0.31$. Figure \ref{fig:gas_fraction} shows
the mass fraction of cold, cool, warm, and hot gas as a function of distance from the cluster center. 

As expected, hot gas dominates (> 90\% of the mass) in the inner cluster regions. At $r=2.5\mathrm{R}_{200}$ from the cluster center, there is
an equal amount of mass in the hot phase and the combined cold, cool, and warm gas phases. At $r=3\mathrm{R}_{200}$ from the cluster center, 
the warm gas comprises $\sim 45 \%$ of the total gas mass while
the hot and  cool gas phases each comprise $\sim 35\%$ and $\sim 15\%$ of the gas mass, respectively. At these radii, roughly 80\% of the 
gas mass is in the cool-warm phase that could potentially be probed by UV observations. Even at the cluster outskirts, cold gas takes up
less than 10\% of the total gas mass. Although we don't expect the mass fraction of cold gas to be higher than the cool or warm phases, 
the fraction of cold gas in our simulation may lowered by limitations in resolution. For example, because the simulation cannot resolve very dense regions of very cold gas, the density and temperature thresholds for star formation are relatively low and high, respectively in {\sc RomulusC}.

These results are broadly consistent with the gas mass fractions in the Virgo-sized cluster simulation presented by \citet{Emerick:2015}. 
Both simulations predict that hot gas dominates in the inner ICM, that the temperature profile gradually drops with 
distance from the cluster, and that the cluster outskirts have substantial reservoirs of cool-warm gas. 
The greatest difference is that our simulations predict a larger fraction of cool and warm gas at the cluster outskirts.

\subsection{Covering fraction}
\label{sec:covFrac}

The presence of gas in the $\mathrm{T}<10^6$ K temperature regime manifests in observable spectral signatures, namely absorption lines of \hone\ and metal ions.  A fundamental metric of any absorption-line survey is the covering fraction, or detection rate, of a given species, defined as the number of sightlines in which that species is observed with some column density (or equivalent width) above a given threshold.  To provide a quantitative measure against which our simulation may be tested, we make predictions for the covering fractions of \hone\, \cfour, and \osix\ absorption
given a number of detection thresholds throughout the cluster. These three ions span the temperature range of the cool-warm gas phases and have strong UV transitions observable with \emph{HST/COS} at low redshifts.  With the medium-resolution \emph{FUV COS} gratings, G130M and G160M, the \hone\ \lya\ line is observable at $z\lesssim0.5$, while the \osix\ doublet ($\lambda$ $\lambda$ 1031.9, 1037.6 \AA) shifts into the sensitive wavelength regime of HST at $z\gtrsim0.11$.  The \cfour\ doublet ($\lambda$ $\lambda$ 1548.2, 1550.8 \AA) is observable down to $z=0$ but exits the G160M bandpass at $z>0.16$ 

Figure \ref{fig:cfrac_threshold} shows the covering fractions of \hone, \cfour, and \osix\ as a function of impact parameter measured from the cluster center.  The curves are colored according to various column density thresholds, inevitably showing higher covering fractions for more sensitive thresholds.  The darker curves correspond to stronger absorption features, which may be more easily detected in data with lower signal-to-noise ratios (S/N).  As we discuss in more detail in \S\ref{sec:prospects}, S/N$\sim10$ spectra are typical for COS, which correspond roughly to $3\sigma$ detection limits of $10^{13}~\cmt$ for \hone\ and $3 \times 10^{13}~\cmt$ for \cfour\ and \osix.  

Although the covering fractions of all ions decline with impact parameter, the \hone\ covering fraction declines more slowly than the covering fractions of \osix\ and \cfour.  The gas density, temperature, and metallicity all impact the resulting ion abundances, and the various ion tracers constrain the physical conditions of the gas.  With the nominal column density limits quoted above, one should detect \hone\ in a majority of sightlines within 1.5 Mpc of the cluster center, \osix\ in $\sim15\%$ of sightlines, and \cfour\ in $<10\%$.  \osix\ has an ionization potential (114 eV) that is more than double that of \cfour\ (48 eV), and it is likely that even the warm ($\mathrm{T}>10^5$K) gas present in the ICM is too hot for carbon to survive in this ionization state; \cfour\ peaks under collisional ionization at $\sim 10^{5.1}$ K \citep{Gnat:2007fk}.  We measured the statistics for other, lower ionization species observable with COS, such as \cthree\ and \sithree, showed similarly small covering fractions.  Interestingly, the covering fractions suggest that it is possible to detect the presence of warm and cool gas even at the cluster center. Increasing the column density sensitivity by an order of magnitude would approximately double the expected covering fractions throughout the ICM, and we discuss prospects for increased sensitivity in \S\ref{sec:prospects}.

In a survey using several sightlines probing the Virgo cluster, \citet{Yoon:2012yu} report an \emph{increase} in \hone\ \lya\ covering fraction between $<1.5$ Mpc ($0.60^{+0.16}_{-0.13}$) and $>1.5$ Mpc ($1.00_{-0.14}$) for absorbers with N(\hone) $> 10^{13.3}~\cmt$ and a covering fraction of unity for N(\hone) $> 10^{13.1}~\cmt$ absorbers.  \citet{Yoon:2012yu} attribute this to substructure within Virgo and on its outskirts.  \citet{Yoon:2017aa} conducted a similar study on the Coma Cluster and found a similar covering fraction, albeit without such a difference between inner and outer radii.  In a sample of X-ray bright clusters, \citet{Burchett:2018aa} detected \hone~ absorbers in 4/5 clusters. In addition, their data show a hint of increased absorption in the cluster outskirts.  \citet{Muzahid:2017lr} detected very strong \hone\ absorbers on the outskirts of three SZ-selected clusters at $R > 1.6 R_{500}$, although at higher redshift ($z>0.4$). Various factors such as cluster dynamical state, redshift, and mass may indeed impact the presence of these absorption-line indicators, but greater amounts of cool gas may reside in the cluster outskirts than our simulations indicate.

\begin{figure}
	\includegraphics[width=0.5\textwidth]{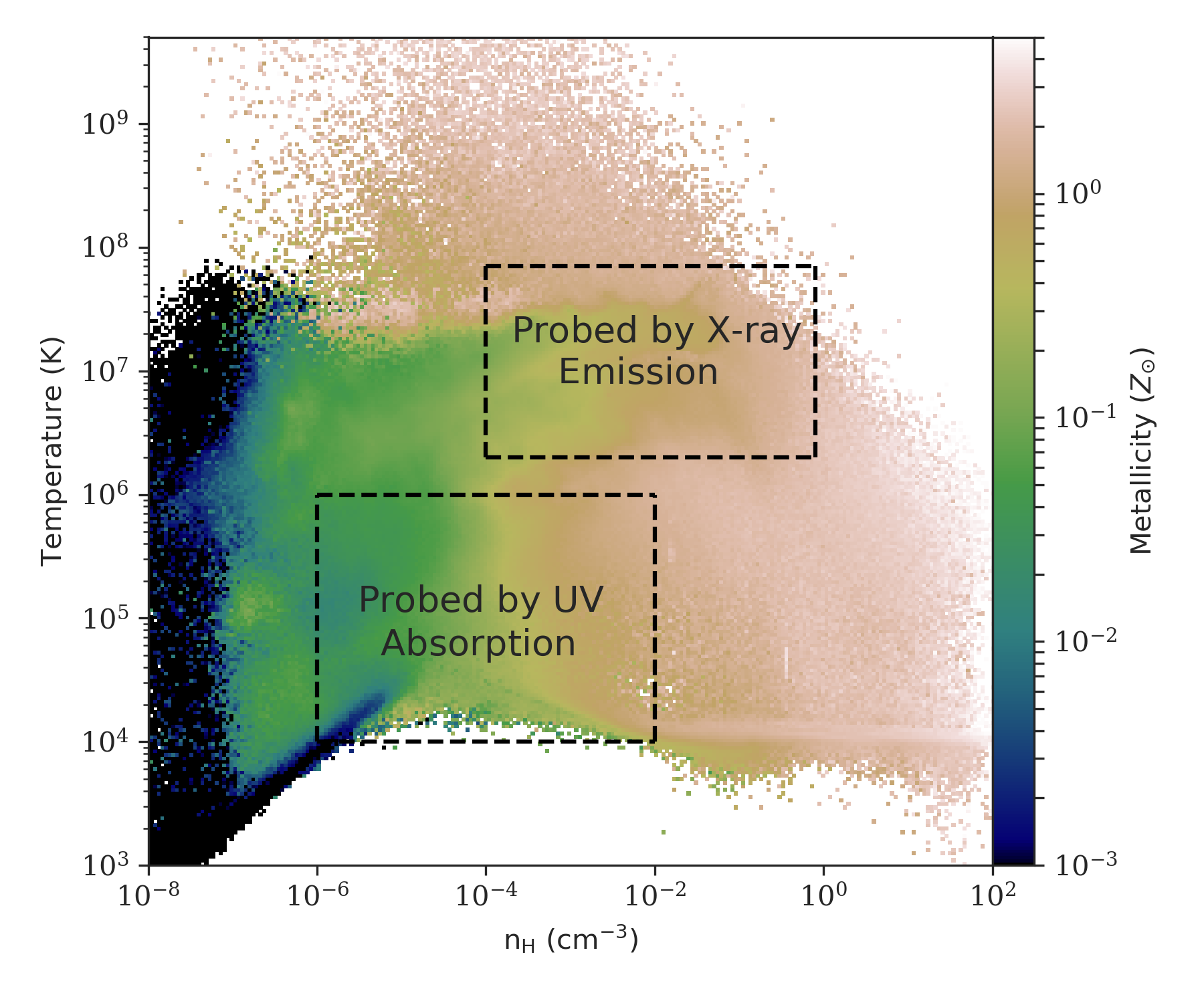}
    \caption{The distribution of the mass-weighted, average metallicity as a function of the temperature and hydrogen number density of the gas. The plot data was taken from a snapshot of RomulusC at $z = 0.31$, within 3$\mathrm{R}_{200}$ of the cluster center. Gas probed by X-rays exists in a disjoint temperature and density phase from gas probed by UV-lines.  The metallicity inferred from X-ray observations is likely to be higher and more uniform than the metallicity probed by UV lines. 
     \label{fig:metallicity}}
\end{figure}
 \begin{figure}
	\includegraphics[width=0.5\textwidth]{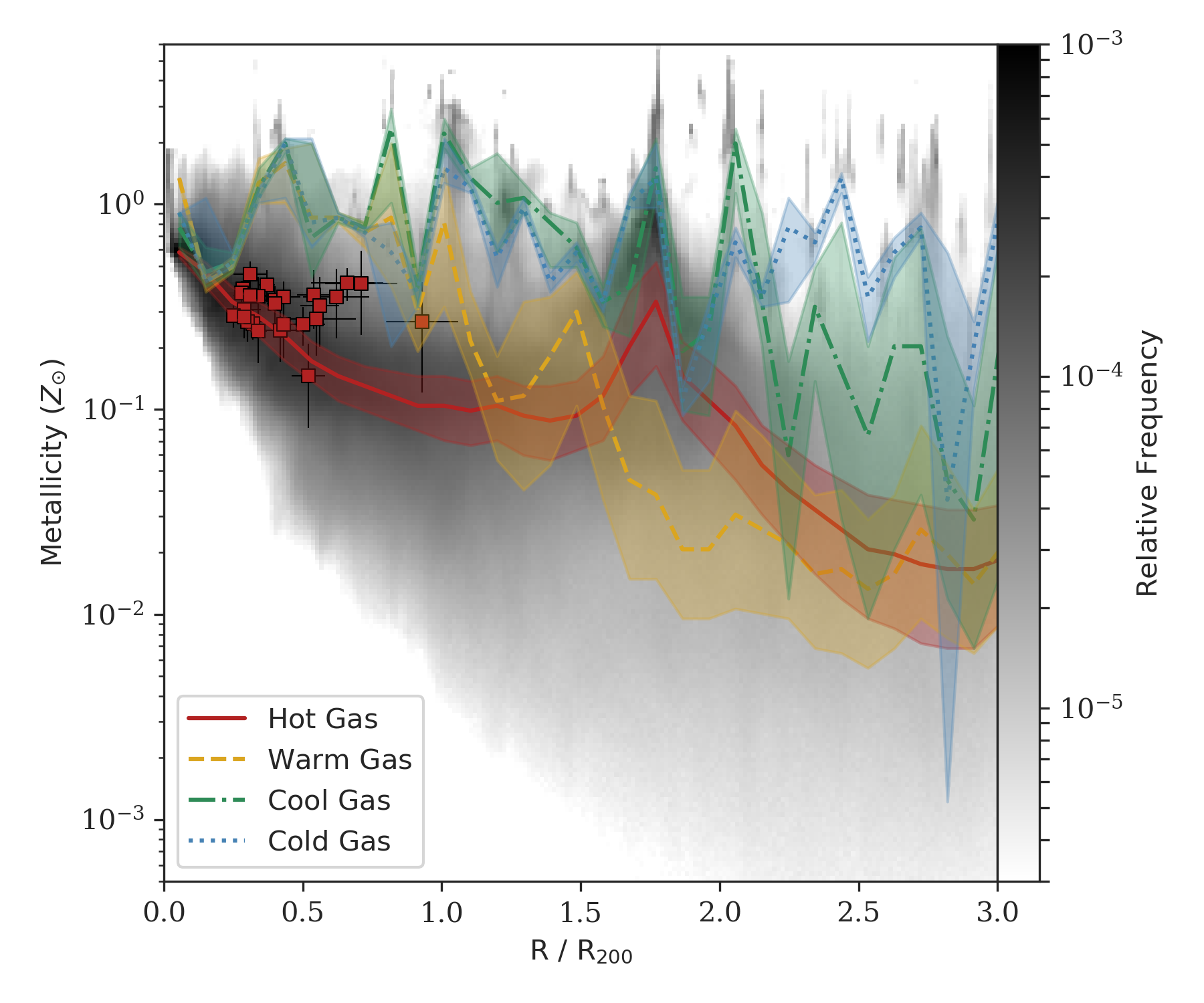}
    \caption{The 3D radial profiles of the median metallicity in cold ($\mathrm{T}< 10^4$K), cool ($10^4 < \mathrm{T} < 10^5$K), warm ($10^5 < \mathrm{T} < 10^6$K), and hot ($\mathrm{T} < 10^6$K) gas phases at $z = 0.31$. The shaded regions represent the 68.8\% confidence intervals. The gray pixels in the background show the normalized mass distribution of all of the particles in the simulation within 3R$_{200}$ of the cluster center. 
    The red squares denote the X-ray derived metallicities from \citet{Urban:2017}. \label{fig:metallicity_radius}}
\end{figure}
 \begin{figure*}
\begin{center}
	\includegraphics[width=\textwidth]{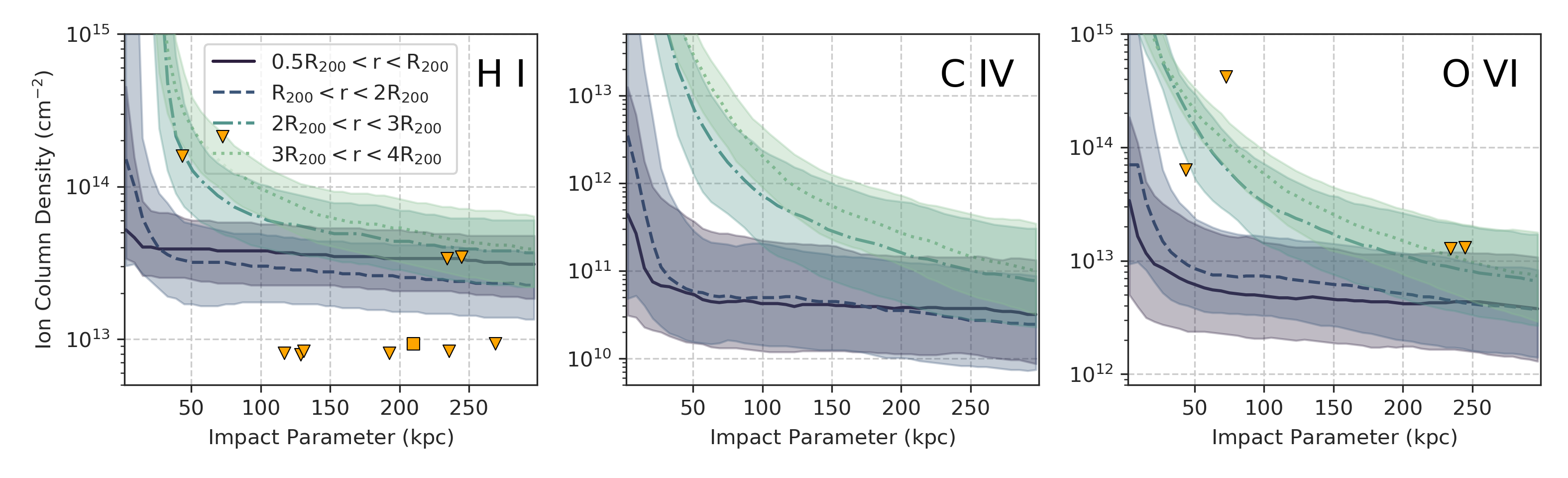}
    \caption{\textit{Left to right:} The column density of \hone, \cfour, and \osix\ as a function of impact parameter for all galaxies with stellar masses $M_* > 10^9M_{\odot}$ within 4 $R_{200}$ of the cluster center. The lines plot the median column density profiles of galaxies within certain bins of distance from the cluster center as indicated in the legend. The shaded regions indicate the 68.8\% confidence interval of the distribution.  The orange shapes show the observed detections (square) and upper limits (triangles) of \hone\ and \osix\ in cluster galaxies from \citet{Burchett:2018aa}.}    
    \label{fig:column_halo}
\end{center}
\end{figure*}

\subsection{Distribution of metals}
As stated above, the diverse range of ICM structures probed by UV absorption should detect gas with different metallicities than
the gas typically observed through X-ray emission.
Figure \ref{fig:metallicity} shows the distribution of gas temperature and density around 3$\mathrm{R}_{200}$ around the 
cluster center. The pixels are colored by the mass-weighted average metallicity of the gas that lies in each temperature-density bin.
The black square boxes roughly outline the density and temperature ranges that are probed by X-ray and UV observations. 

The metallicity of gas probed by X-rays is roughly uniformly $0.3 Z_{\odot}$ with a small dispersion. At the high temperature end ($\mathrm{T} > 3\times10^7$K),
the average metallicity of the gas is consistently above solar metallicity. 
This is likely dominated by hot, dense, and metal-rich 
supernova or AGN ejecta throughout the simulation. In contrast, the low-density region at lower temperatures exhibits lower metallicities, likely tracing the 
diffuse cluster outskirts that are, on average, less polluted with metals than the ICM near the cluster center (see Figure \ref{fig:multipanel}). Overall, the bulk
of the metals traced by X-ray observations belong to the hot, dense ICM. The X-ray emission is strongest near the cluster
center, where the metals have been thoroughly mixed in the turbulent ICM.

Gas traced by the UV has a significantly larger variation in its average metallicity. In this regime, the metallicity is also a tracer
of the gas density. Intuitively, this wide range in metallicity traces gas being stripped from galaxies within and around the cluster.
For example, in Figure \ref{fig:multipanel}, the filamentary structures in the projected metallicity often trace the filamentary structures in the temperature and density. The ISM and CGM of galaxies are stripped of their high-metallicity gas as they move
through the ICM. Over time, these metals mix with the ambient ICM. With the sensitivity to a much wider range in densities, UV observations can better measure the inhomogeneity of ICM the metallicity.

In Figure \ref{fig:metallicity_radius}, we further quantify the distribution of metals in each of the different gas phases (red, orange, green, and blue for hot, warm, cool and cold, respectively) by plotting the median metallicity profiles of each temperature phase. The shaded areas show the 68.8\% confidence interval of each gas distribution.
The grey pixels show the normalized distribution of all gas (without temperature cuts) within 3 Mpc of the cluster center. 
The red squares show X-ray measured metallicities from \citet{Urban:2017}.

This radial metallicity profile highlights the range of metallicities in the multiphase gas predicted in Figure \ref{fig:metallicity}.
For example, consider the inner Mpc around the cluster center, which is composed of mostly hot gas (see Figure \ref{fig:gas_fraction}). 
Although the observed metallicities are roughly constant at 
$0.3Z_{\odot}$, our simulation predicts that the spread in metallicities (considering all phases) spans significantly more than an order of magnitude. This spread is even larger at the cluster outskirts, reflecting the mixture of multiphase gas comprising the ISM, CGM and ICM with a wide range of temperatures and metallicities.

Furthermore, the different gas phases have varied metallicity profiles. The median metallicity of the hot gas phase is highest at the cluster center
and decreases with radius. Although there is still a substantial mass of hot gas at the cluster outskirts, the metals from stripped
CGM are not as well mixed as that near the cluster center. We predict that (by mass) the median metallicity of hot gas beyond 0.5 R$_{\rm 200}$ is actually below that reflected by X-ray observations. The metallicity profiles of the cool and cold phases have higher median metallicity values than the hot phase at nearly all radii, because of the metal-rich gas in the cool CGM and ISM of galaxies within the cluster. However, the metallicity profile of the warm gas falls off more rapidly than that of the cool or cold gas, and the warm gas metallicity is more representative of the global median at $>1.5$ R$_{\rm 200}$. Also beyond 1.5 R$_{\rm 200}$, the
median profiles of the cool and cold gas show large variations due to the stochasticity in the distribution of galaxies and thus track relatively well with the infalling galaxies' CGM metal contents. 

While the lack of metals present beyond $0.5\mathrm{R}_{200}$ may indicate that {\sc RomulusC} fails to transport metals as effectively as it should, we stress that these results are sensitive to the redshift at which we sample the simulation. At redshifts more consistent with observations ($z<0.2$), the cluster outskirts are more metal-enriched. For the sake of consistency, we limit our analysis to $z=0.31$ and will explore the time evolution of the ICM in future work. We also stress that no attempt was made to mimic X-ray observations beyond selecting for hot gas. As X-ray observations are more sensitive to denser gas, they may be biased toward higher metallicity relative to the median presented in Figure \ref{fig:metallicity_radius}.

\begin{figure*}
    \includegraphics[width=\textwidth]{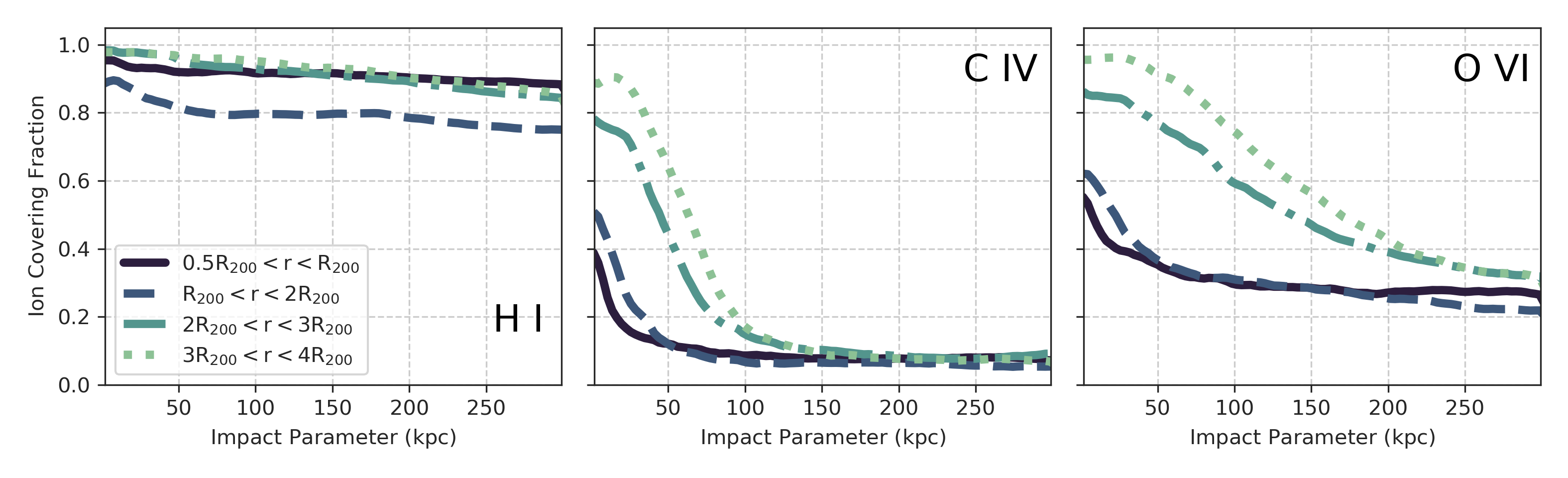}
    \caption{The covering fractions of \hone, \cfour, and \osix\ as a function of halo-centric
        impact parameter for all galaxies with $M_* > 10^9 M_{\odot}$ at z = 0.31. The different lines represent halos
        within the indicated distance bins from the cluster center. Covering fractions measure the fraction column densities above some threshold column density.  We assume column density thresholds of $10^{13}\mathrm{cm}^{-2}$ for \hone, $10^{13.1}\mathrm{cm}^{-2}$ for \cfour, and $10^{13.3}\mathrm{cm}^{-2}$ for \osix. The covering fraction of all ions is lower for galaxies nearer to the cluster center.}
    \label{fig:cfrac_halo}
\end{figure*}

\begin{figure}
	\includegraphics[width=0.5\textwidth]{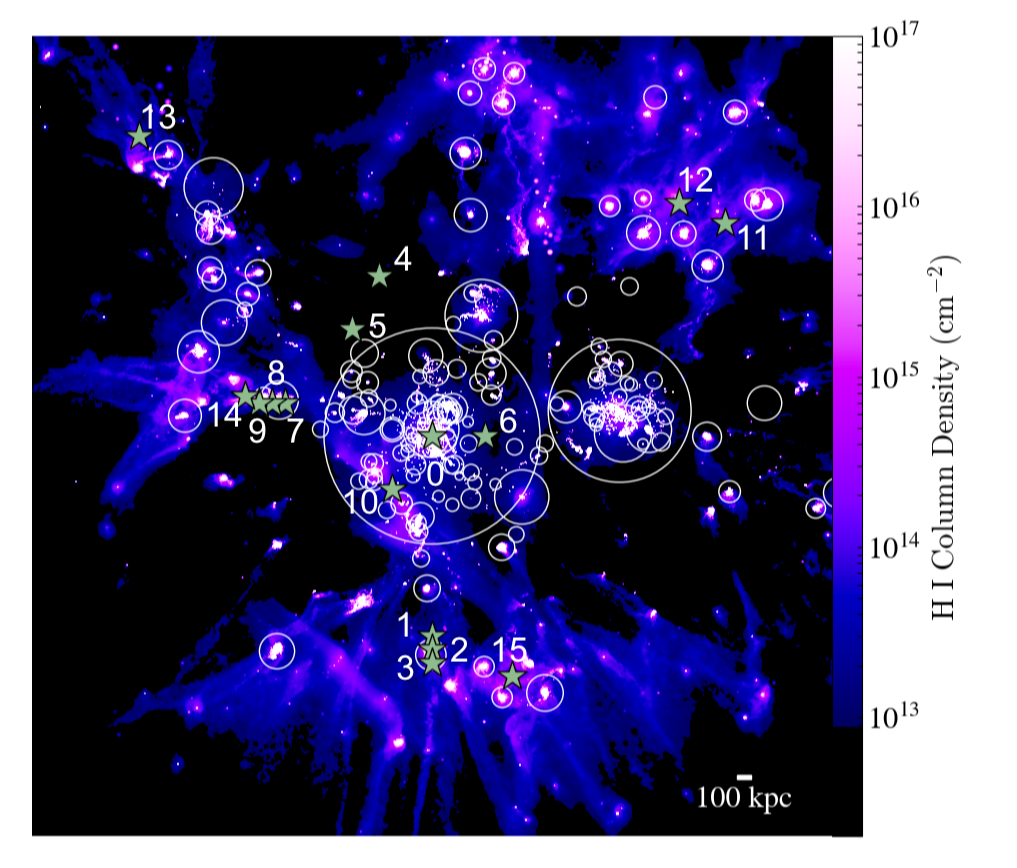}
    \caption{The \osix\ column density overplotted with locations of sightlines for which we extracted synthetic spectra.  The virial radii of cluster galaxies with stellar masses of $M_{*} > 10^{9} M_{\odot}$ are indicated by circles (centered on each galaxy). }
    \label{fig:sightlineMap}
\end{figure}

\begin{figure*}
	\includegraphics[width=\textwidth]{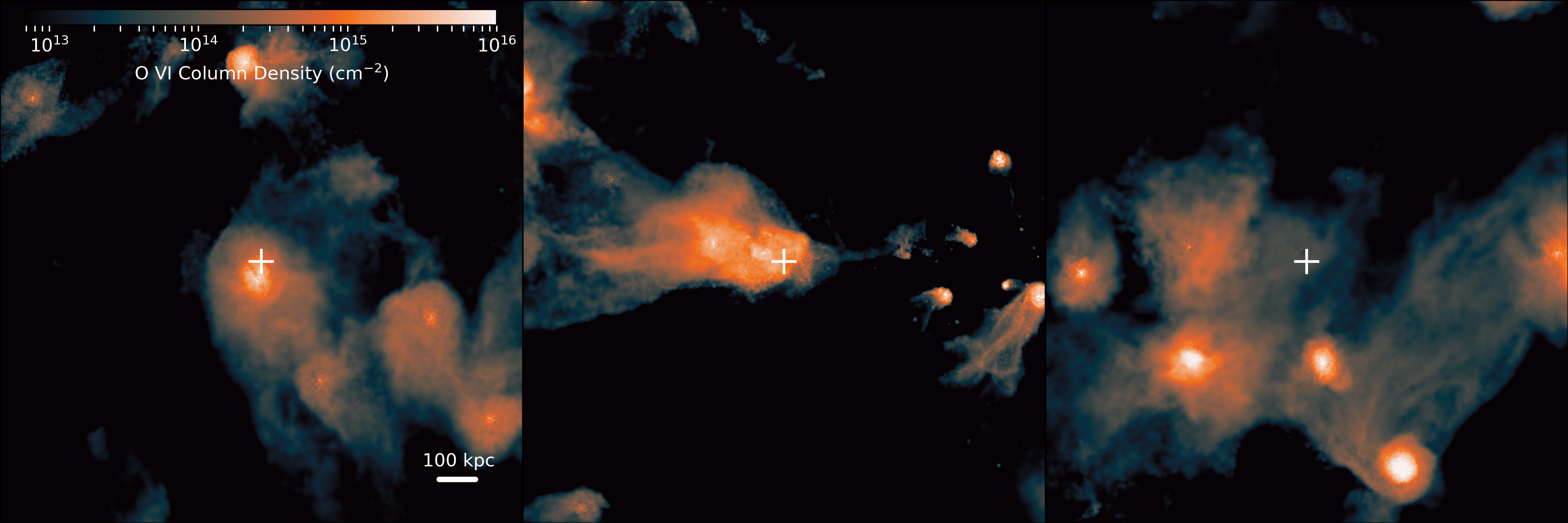}
    \caption{The \hone\ column density around the regions of the sightlines presented in Figure \ref{fig:spectra}. 
        Each sightline was chosen at an arbitrary location, in-between galaxies at the $z = 0.31$ snapshot of {\sc RomulusC}. 
        It is not required that a sightline directly intersect the CGM of a galaxy for there to be detectable amounts of \osix\ in the ICM.
        }
    \label{fig:spectraLocations}
\end{figure*}

\begin{figure*}
	\includegraphics[width=0.3\textwidth]{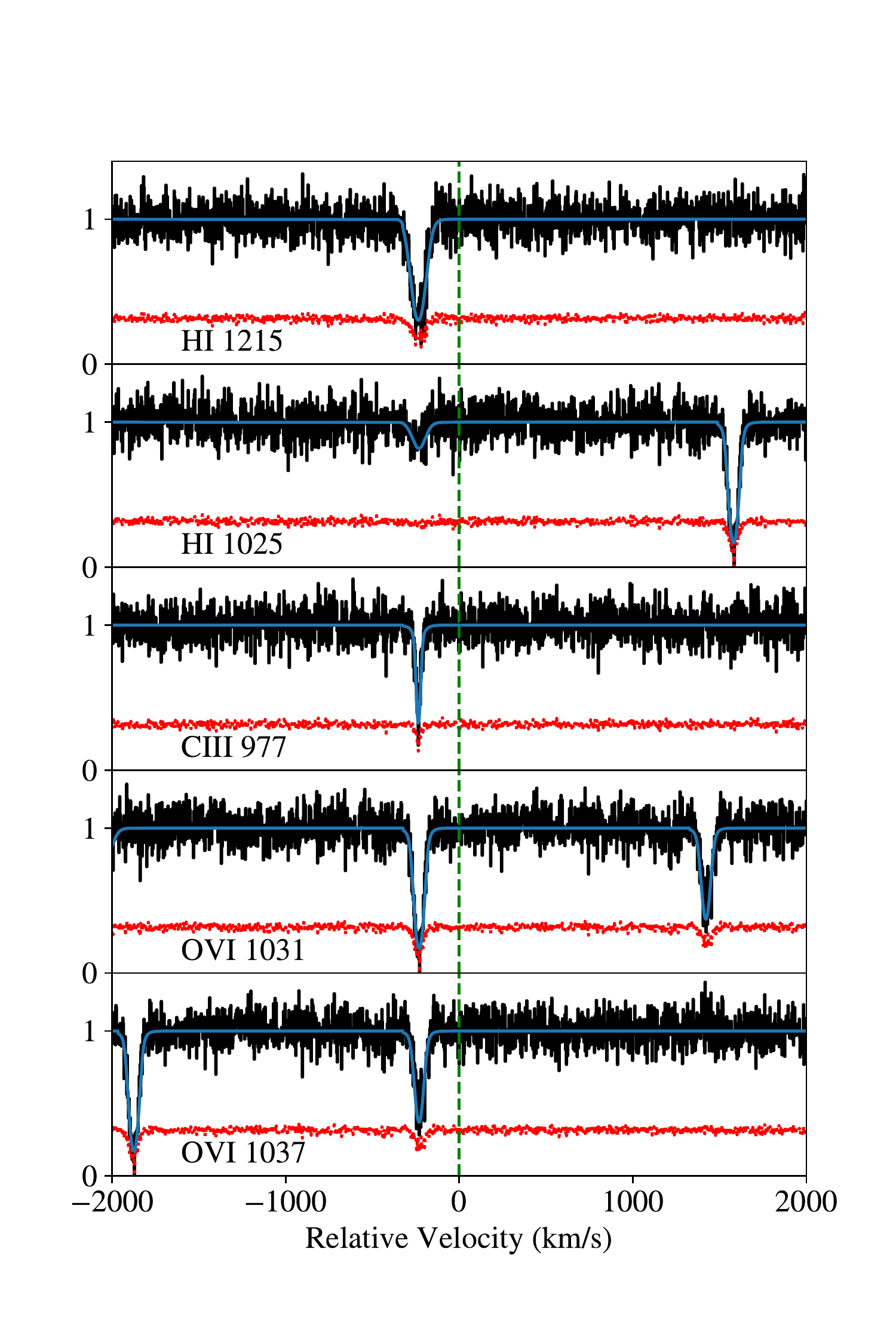} 
  	\includegraphics[width=0.3\textwidth]{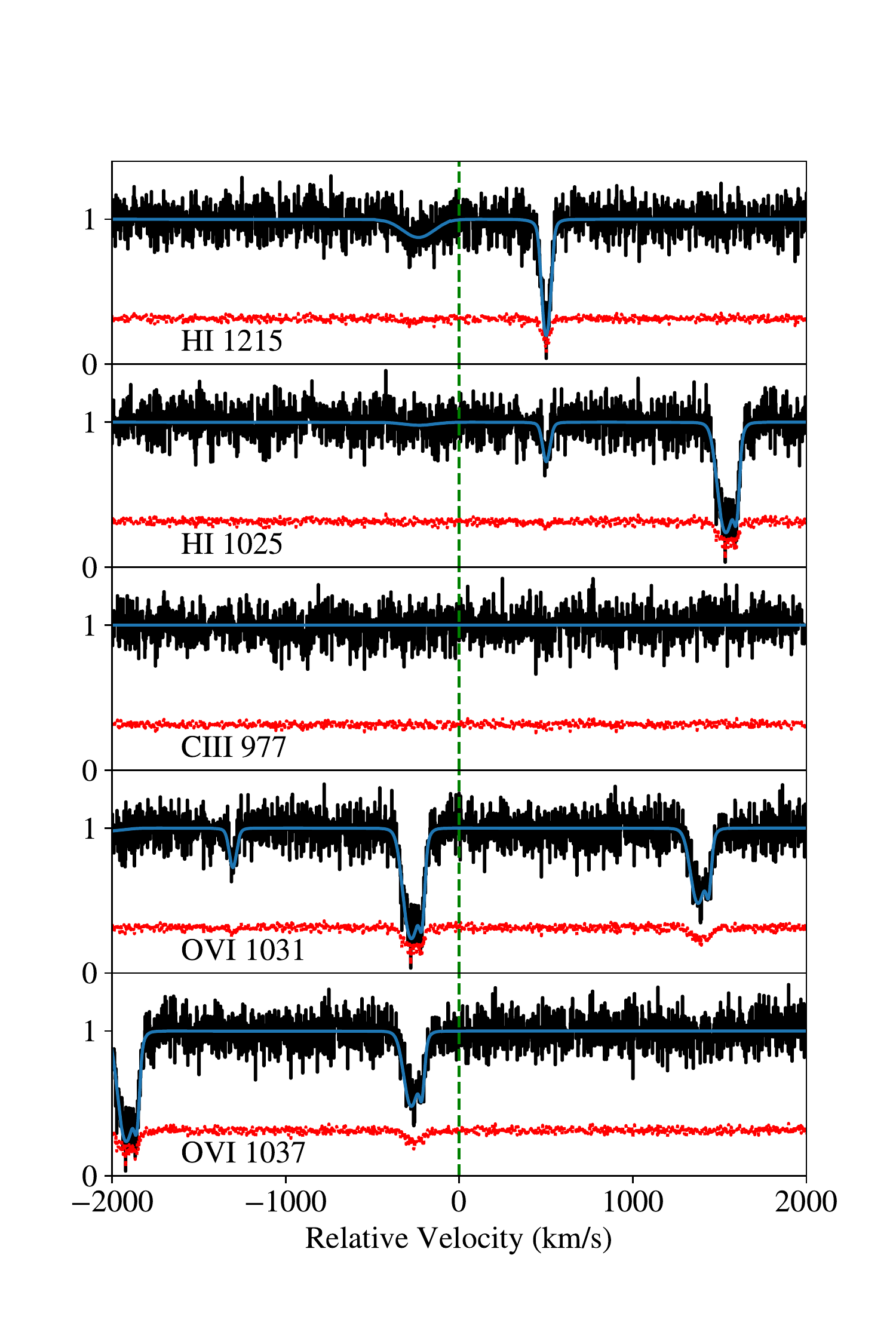} 
	\includegraphics[width=0.3\textwidth]{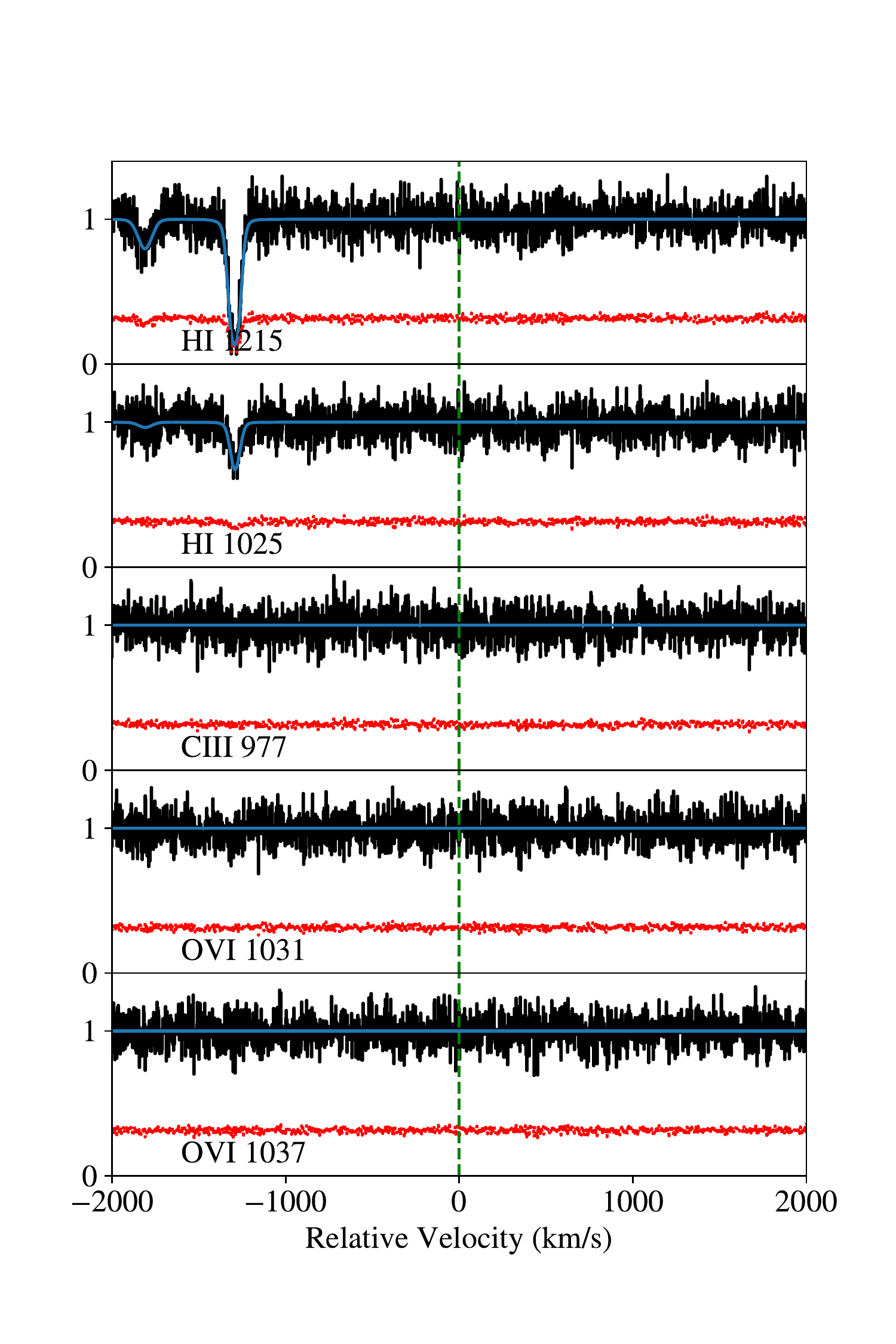}
    \caption{Synthetic spectra of sightlines generated using {\sc Trident} through the {\sc RomulusC} cluster. From left to right, the panels correspond to Sightlines 2, 7, and 12 marked in Figure \ref{fig:spectraLocations}. Each subpanel shows absorption (or lack thereof) for the transition labelled in its lower right corner, relative to $z=0.31$. The blue curves mark Voigt profile fits to the visually identified absorption components.  Note that the apparent absorption at $>1000$ and $<1000$ km/s in the Sightline 2 and 7 \hone\ 1025, \osix\ 1031, and \osix\ 1038 panels are actually \osix\ 1031, \osix\ 1037, and \osix\ 1031 lines, respectively, due to the large velocity range plotted.  These spectra show the diversity of absorbers predicted by our simulation, including broad and narrow \hone\ components with and without corresponding \osix\ and \cthree.}
    \label{fig:spectra}
\end{figure*}

\section{The CGM of RomulusC Cluster Galaxies}
\label{sec:CGMinClusterGals}

The hot and dense ICM plays an important role in the evolution of cluster galaxies. For example, galaxies moving through 
the ICM often experience strong ram-pressure forces that strip them of their gas and quench star formation. Although this effect appears in many modern simulations \citep[e.g.][]{Bahe:2017aa,Zinger:2018aa,Tremmel:2019}, the quenching timescales differ depending on whether the ISM is stripped directly or indirectly deprived of replenishing fuel from a CGM that has been stripped early upon infall.  We explore the effects of the cluster environment upon the observable ion abundances in sightlines probing the CGM of cluster galaxies at various radii. 

In Figure \ref{fig:column_halo}, we compare the ion column densities of \hone, \cfour, and \osix\ as a 
function of impact parameter measured from the centers of cluster-galaxy halos. Each line shows the median 
column density for all galaxies with stellar masses of $M_{*} > 10^9 M_{\odot}$ within 
the bins of projected distance from the cluster center indicated in the legend. Each shaded regions contain the 68\% confidence 
interval of the distribution in each radial bin. The orange markers show the observed detections (square)
and upper limits (triangle) of \hone\ and \osix\ in cluster galaxies from \citet{Burchett:2018aa}. 
Although \cfour\ does not have comparable observations, we include
its column density profile as this ion is observable with HST/COS at the lowest redshifts. Note that we do not include predictions for galaxies within $0.5\mathrm{R}_{200}$. We find that, in this regime, the `detected' ions (especially \hone) are dominated by the ambient ICM and are not necessarily signatures of the CGM of individual galaxies.

Our simulation predictions are broadly consistent with the limited existing CGM \osix\ observations.  However, the \hone\ column densities we predict at impact parameters $>100$ kpc  generally exceed the observational data.  A preliminary check of lower-redshift (z = 0.1, 0.2) simulation snapshots show lower \hone\ CGM column densities more consistent with the observations but that are still high. We will explore this redshift evolution further in future work.  Nevertheless, this inconsistency may also reflect some aspect of the  galaxy-CGM-ICM interaction that is not accurately reproduced by our simulation.  Larger samples of observational data will prove invaluable to constraining the simulated physics.

Notably, we predict a decrease in CGM column densities at impact parameters $\lesssim200$ kpc for galaxies with projected clustocentric distances of $<2\mathrm{R}_{200}$. Below distances of $2\mathrm{R}_{200}$ there is little evolution of the column density profiles beyond the very innermost ($\lesssim 25$ kpc) impact parameters. The larger clustocentric bins show progressively decreasing profiles from $>3\mathrm{R}_{200}$ to $2< \mathrm{r}/\mathrm{R}_{200}<3$ before the more dramatic decline at $2\mathrm{R}_{200}$. Thus, the majority of the CGM is stripped while galaxies are still quite far from the cluster center; however, the innermost CGM requires much greater ram pressure to become stripped, as it is the most bound. It is interesting to note that, at large impact parameters, the column densities we predict for \osix\ are lower than many of the observations from the COS-Halos isolated galaxy sample \citep{Tumlinson:2013}.  Simulations of isolated galaxies using the same sub-grid physics models as \textsc{RomulusC} recover the COS-Halos \osix\ column densities quite well \citep{Sanchez:2018}, indicating that the cluster environment is inducing strong effects on the CGM of galaxies even out to clustocentric distances of $4\mathrm{R}_{200}$. We defer a more in-depth comparison between simulated cluster and isolated galaxies as well as observations thereof to a future work.

In Figure \ref{fig:cfrac_halo}, we show the covering fractions of \hone, \cfour, and \osix\ as a function of impact parameter for the same galaxies considered in Figure \ref{fig:cfrac_halo}.
Each line represents the combined covering fraction of all galaxies within each distance bin calculated for the observationally motivated column density thresholds of $10^{13}$, $10^{13.1}$, and $10^{13.3}~\cmt$ for \hone, \cfour, and \osix, respectively. The covering fraction of \cfour\ is low at large impact parameters even for galaxies in the outskirts of the cluster. At smaller impact parameters, \cfour\ becomes depleted for galaxies within $2\mathrm{R}_{200}$ from the cluster center. A similar effect is seen for \osix, although the covering fraction is generally higher than that of \cfour\ at all impact parameters and the profile steepens significantly within $2\mathrm{R}_{200}$. The covering fractions of \hone\ decline very modestly with impact parameter, and the evolution with clustocentric distance is much greater for the metal ion diagnostics \osix\ and \cfour. 
As a general trend, galaxies farther from the cluster center tend to retain more of their CGM, resulting in higher covering
fractions of \cfour\ and \osix;
ram-pressure increases closer to the cluster center where the density of the ICM is higher.

\section{The physical conditions of cool and warm gas in the ICM from synthetic spectra}
\label{sec:specAnalysis}

Herein, we discuss several facets of a multiphase ICM, which in turn have observable signatures, particularly in the UV regime for $\mathrm{T}<10^6$K gas.  In this subsection, we seek to 1) further assess the observability of this material and 2) relate the observable signatures back to the physical conditions of the gas in the {\sc RomulusC} simulation.  Our intent is to provide a proof-of-concept level of analysis while sampling various regions throughout the cluster, as would be probed in an actual absorption-line spectroscopic survey using sightlines piercing galaxy clusters. 

Using Trident \citep[][discussed in \S\ref{sec:synthSpec}]{Hummels:2017}, we extracted sixteen sightlines through the {\sc RomulusC} cluster; the sightline locations are indicated in Figure~\ref{fig:sightlineMap}.  The sightlines were chosen to pass through ambient ICM regions far away from infalling galaxies, the immediate CGM of cluster galaxies, and detritus stripped as galaxies fall into the cluster.  The virial radii of subhalos within the cluster are marked with circles in Figure \ref{fig:sightlineMap}.  We then analyzed the resulting synthetic spectra as an observer might analyze such data probing an actual cluster obtained with \emph{HST/COS}, the only currently available instrument that is capable of producing a similar dataset (see discussion in \S\ref{sec:prospects}).  

Three examples of spectra we extracted are shown in Figure \ref{fig:spectra}; in each plot, the individual panels show the would-be locations of \hone\ \lya\ and \lyb, the \osix\ $\lambda$ 1031, 1037\AA\ doublet, and \cthree $\lambda$ 977\AA\ within 2000 km/s of the cluster redshift.  For each spectrum, we searched by-eye for absorption signatures of these four lines, marking the approximate velocity of each component.  In total, we visually identified 20 components of \hone, \osix, or \cthree\ in our synthetic spectra sample. 

Two key factors hinder detecting weaker lines: the S/N of the data and line broadening due to thermal and non-thermal motions.  Thus, the wavelength coverage of multiple transitions enables one to corroborate identifications of features using the presence of other lines.  For example, a strong \lya\ line should be accompanied by a weaker (the ratio set by atomic physics) \lyb\ line at the same velocity.  Similarly, albeit with caution, one can corroborate lines of one species (e.g., \osix) with absorption by another (e.g., \hone).  While this practice can lead to misidentifications if applied improperly, it is occasionally necessary in cases with weak features or blending due to lines from other redshifts.  Our dataset analyzed here (and as would be analyzed from an actual survey) presents a number of such challenges, as the special cluster environment contains an excess of hotter gas relative to random CGM/IGM regions; the increased gas temperature thermally broadens \hone\ lines substantially.  

We highlight one particularly illustrative case that underscores the importance of wavelength coverage in this analysis, shown in the middle panel of Figure \ref{fig:spectra} at $v\sim-250$ km/s.  While no \lyb\ is detected, a broad \lya\ feature is barely apparent.  However, the \osix\ doublet is well detected at this velocity, greatly reinforcing the broad \lya\ identification.  The width of the \lya\ line then enables a direct constraint on the gas temperature.  Furthermore, without the \lya\ coverage, one may identify the \osix\ doublet as a spurious superposition of two \lya\ lines at lower redshifts.  While the \osix\ and \lyb\ lines would be covered in a COS observation with only the G130M grating, observations with the G160M would be required to cover the observed wavelength of \lya.  The symbiotic complementarity of information gleaned from both gratings is essential to both discovering this gas system and understanding its physical conditions.

Upon identifying the absorption features in the synthetic spectra, we fitted Voigt profiles to each feature using the $Veeper$ software originally developed for the analysis of \citet{Burchett:2015rf}.  Measurements for all detected \hone\ and \osix\ lines as well as information about their respective sightlines are provided in Table \ref{tab:sampleTable}.  Notably, we detect several broad \lya\ lines, with seven components having Doppler parameter values $b > 40$ km/s, tracing warm gas at $\mathrm{T}>10^5$K (assuming thermal broadening).  To compare this synthetic sample with previous survey measurements of broad \lya\ absorbers, we perform an approximate conversion of absorber line density as measured through our simulation to the line density metric often reported in absorption-line surveys $d\mathcal{N}/dz$, or number of absorbers per unit redshift.  Our sightlines extend 5 Mpc through the {\sc RomulusC} simulation, which corresponds to the luminosity distance difference over $\Delta z \sim 0.0008$ at $z=0.31$ (assuming cosmological parameters from \citet{Planck-Collaboration:2016eq}).  Thus, over the 16 sightlines we have analyzed, our broad \lya\ density would translate to $d\mathcal{N}/dz \sim 540$.  Compared with the broad \lya\ $d\mathcal{N}/dz \sim 25$ from \citet{Lehner:2007lr}, our broad \lya\ sample represents a $>20$x overdensity relative to random IGM sightlines.  Of course, our measurement is heavily biased, as all of our sightlines fall within the cluster environment, and the incidence of broad \lya\ absorbers should increase in such overdense regions \citep[e.g.,][]{Tepper-Garcia:2012aa}.  Perhaps a more fair comparison is with the survey by \citet{Tejos:2016qv}, who examined the line density near putative intracluster filaments and find a broad \lya\ $d\mathcal{N}/dz = 108^{+65}_{-43}$ within 3 Mpc of such filaments.  We again emphasize that these comparisons are by no means rigorous, as we furthermore did not choose our sightline locations at random locations throughout the cluster, and the pathlength calculation (denominator of these metric) requires accounting for the sensitivity limits of the spectra, etc.  However, our predictions indicate that a substantial number of broad \lya\ absorbers tracing warm gas, well in excess of the typical IGM or even Cosmic Web filaments, should await discovery within the cluster environment.

\begin{table*}
\caption{Voigt profile fitting results for identified absorption components in }
\centering
\begin{threeparttable}
\centering
\begin{tabular}{cccccccc}
\hline \hline 
Sightline\tnote{a} & Ion & log (N/cm$^{-2}$) & b [km/s] & v\tnote{b}~[km/s] & $\rho_{cl}$\tnote{c}~[kpc] & log (T/K)\tnote{d} & log (N$_{H}$/cm$^{-3}$) \\
\hline 
0 & H I & $17.25 \pm 0.30$ & $  15 \pm    1$ & $-357.7 \pm  1.1$ & 0 & 7.10 & -1.88 \\
0 & H I & $16.70 \pm 0.99$ & $  20 \pm    4$ & $-170.0 \pm  1.8$ & 0 & 7.10 & -1.88 \\
2 & O VI & $14.55 \pm 0.04$ & $  28 \pm    3$ & $-233.6 \pm  1.8$ & 1600 & 6.27 & -4.65 \\
2 & H I & $13.94 \pm 0.04$ & $  47 \pm    5$ & $-234.8 \pm  3.4$ & 1600 & 6.27 & -4.65 \\
3 & O VI & $14.13 \pm 0.06$ & $  36 \pm    7$ & $-258.7 \pm  4.4$ & 1700 & 6.15 & -4.84 \\
3 & H I & $13.62 \pm 0.09$ & $  61 \pm   17$ & $-264.2 \pm 10.5$ & 1700 & 6.15 & -4.84 \\
7 & O VI & $14.03 \pm 0.19$ & $  16 \pm    8$ & $-216.3 \pm  4.0$ & 1128 & 6.49 & -4.34 \\
7 & O VI & $14.59 \pm 0.06$ & $  49 \pm    8$ & $-277.9 \pm  6.9$ & 1128 & 6.49 & -4.34 \\
7 & H I & $13.92 \pm 0.05$ & $  24 \pm    3$ & $500.4 \pm  2.2$ & 1128 & 6.49 & -4.34 \\
7 & H I & $13.35 \pm 0.16$ & $ 121 \pm   55$ & $-233.9 \pm 36.7$ & 1128 & 6.49 & -4.34 \\
8 & H I & $17.40 \pm 0.23$ & $  18 \pm    1$ & $-299.2 \pm  1.2$ & 1226 & 6.45 & -4.47 \\
8 & O VI & $14.64 \pm 0.04$ & $  27 \pm    2$ & $-326.4 \pm  1.6$ & 1226 & 6.45 & -4.47 \\
9 & H I & $13.54 \pm 0.10$ & $  49 \pm   17$ & $-279.8 \pm 10.7$ & 1324 & 6.40 & -4.91 \\
9 & O VI & $14.46 \pm 0.04$ & $  25 \pm    3$ & $-284.3 \pm  1.9$ & 1324 & 6.40 & -4.91 \\
10 & H I & $13.26 \pm 0.11$ & $  27 \pm   10$ & $-534.7 \pm  6.7$ & 500 & 6.73 & -4.26 \\
11 & H I & $13.23 \pm 0.13$ & $  42 \pm   17$ & $-1310.8 \pm 11.6$ & 2720 & 5.43 & -5.38 \\
12 & H I & $14.11 \pm 0.04$ & $  32 \pm    3$ & $-1293.5 \pm  2.0$ & 2547 & 5.49 & -5.31 \\
12 & H I & $13.23 \pm 0.14$ & $  50 \pm   22$ & $-1810.2 \pm 14.7$ & 2547 & 5.49 & -5.31 \\
15 & H I & $13.55 \pm 0.08$ & $  69 \pm   16$ & $-1092.0 \pm 10.9$ & 1897 & 5.79 & -5.34 \\
\hline 
\end{tabular}
\begin{tablenotes}
\item[a] Sightline numbers correspond to those marked in Figure \ref{fig:sightlineMap}.
\item[b] Velocities are expressed relative to the RomulusC snapshot redshift at $z=0.31$.
\item[c] Impact parameter of sightline relative to the cluster center
\item[d] Quantity averaged along sightline
\end{tablenotes}
\label{tab:sampleTable}
\end{threeparttable}
\end{table*}

\section{Prospects of studying the cool and warm gas in Galaxy Clusters}
\label{sec:prospects}
We now turn to using the results from our simulation to observationally constrain the multiphase ICM 
and the CGM of cluster galaxies. In \S\ref{sec:multiphaseICM} we show that a substantial fraction 
of the ICM is composed of cool and warm gas beyond 1.5 $\mathrm{R}_{200}$. This gas can be traced observationally 
through \hone, \osix, and (to a lesser extent) \cfour. In \S\ref{sec:covFrac} and \S\ref{sec:CGMinClusterGals} 
we predict the covering
fractions of these three ions as a function of both clustocentric and halo-centric impact parameters. 
Here, we further examine the feasibility of observing the cool and warm cluster gas with existing and
future UV telescopes. 

The predicted covering fractions depend on both the physical abundance of the ion and the column density sensitivity
limits of the observing instrument. In our calculations, we assume constant column density sensitivities of \emph{HST/COS} set
by the anticipated S/N of the spectra. However, in practice, the sensitivity limits depend on several different factors, 
and a given S/N does not directly map onto one sensitivity limit. For example, warm gas will produce more thermal 
broadening than cool gas, which will result in shallower absorption features and a higher minimum observable column
density even with the same S/N. 
Furthermore, stochastic noise can make some otherwise detectable lines insignificant. Instead, the more 
reliable metric for determining the column density sensitivity of an instrument is the detection significance. 
The detection significance uses the S/N, column density, temperature (assuming thermal broadening), and the
instrument-specific line-spread function to measure the ratio of the line width and its uncertainty, $W / \sigma_W$. 
Lines with a detection significance $\ge 3 \sigma$ are considered to be detectable.
We refer the reader to \citet{Burchett:2018aa} for a detailed description of this calculation. 

\begin{figure}
	\includegraphics[width=\columnwidth]{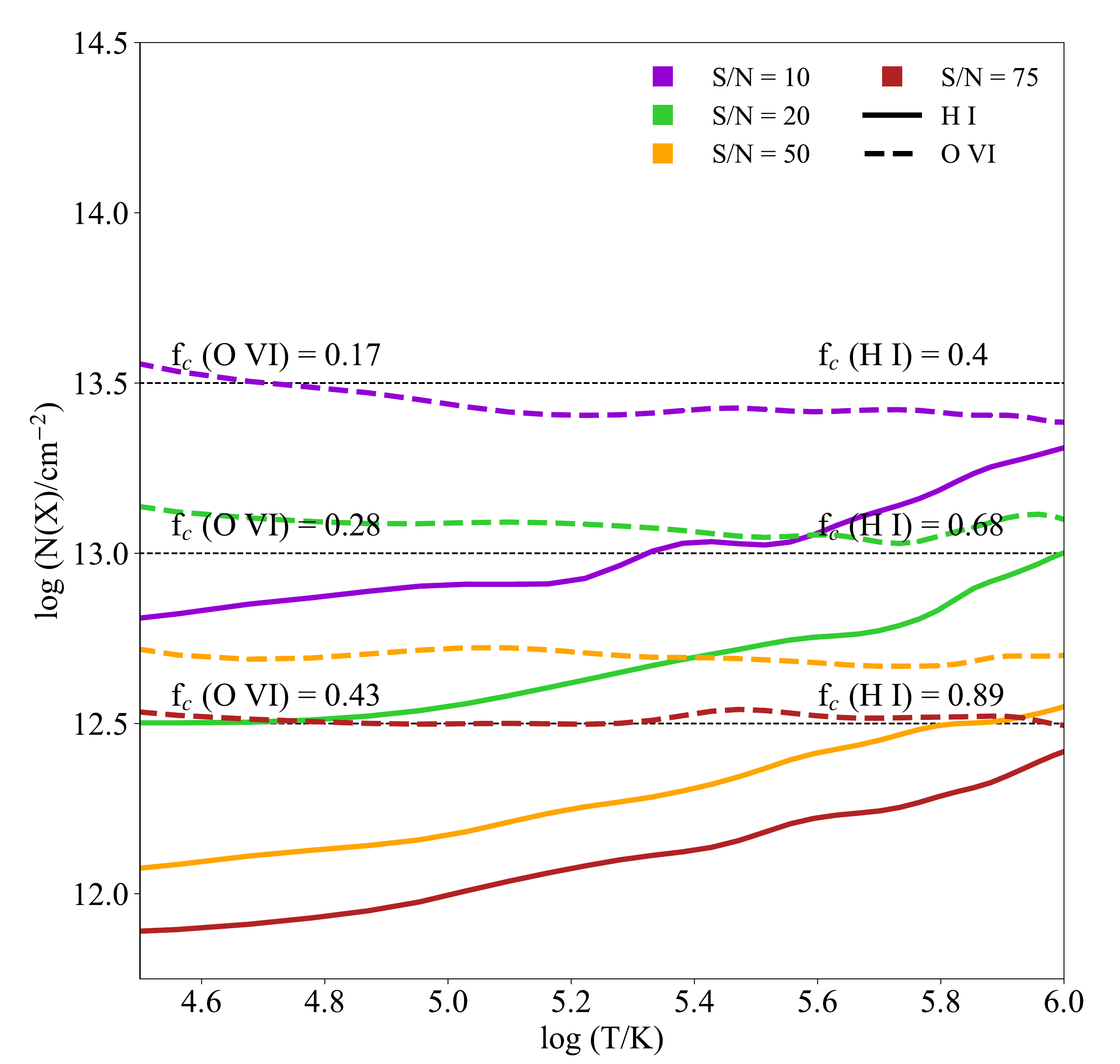}
    \caption{The detectability (at 3$\sigma$) of \hone\ \lya\ (solid) and \osix\ $\lambda$ 1032\AA\ (dashed) lines over a range of column densities and temperatures (via the Doppler $b$ parameter) given several levels of S/N achieved (see color-coding in legend).  Mean covering fractions of \hone\ and \osix\ at three column density thresholds at clustocentric impact parameters $\rho < 2$ Mpc are indicated by horizontal dotted lines.  For example, a S/N = 10 spectrum should return a covering fraction $\sim$17\% for \osix\ across the warm temperature regime but $f_c = 50-70\%$ for \hone\ across this same temperature range.   
    \label{fig:colTempDetect}}
\end{figure}

Figure \ref{fig:colTempDetect} uses the detection significance described above to show the minimum observable
column density of \hone\ and \osix\ as a function of temperature for several different S/N. The dotted horizontal
lines indicate the average covering fractions of \hone\ and \osix\ within $R < 2$ Mpc of the cluster center at 
three different column density thresholds (see Figure \ref{fig:cfrac_threshold}). At each S/N line, 
the minimum observable column density of \hone\ increases with temperature due to the expected thermal broadening. 
In contrast, the minimum observable column density of \osix\ slightly decreases with temperature. This is likely
due to the fact that the collisional equilibrium temperature of \osix\ is $\sim 10^{5.5}$K. For both ions, 
increasing the S/N from 10 to 75 improves the column density sensitivity limit by $\sim$1 dex. 

\begin{figure}
	\includegraphics[width=\columnwidth]{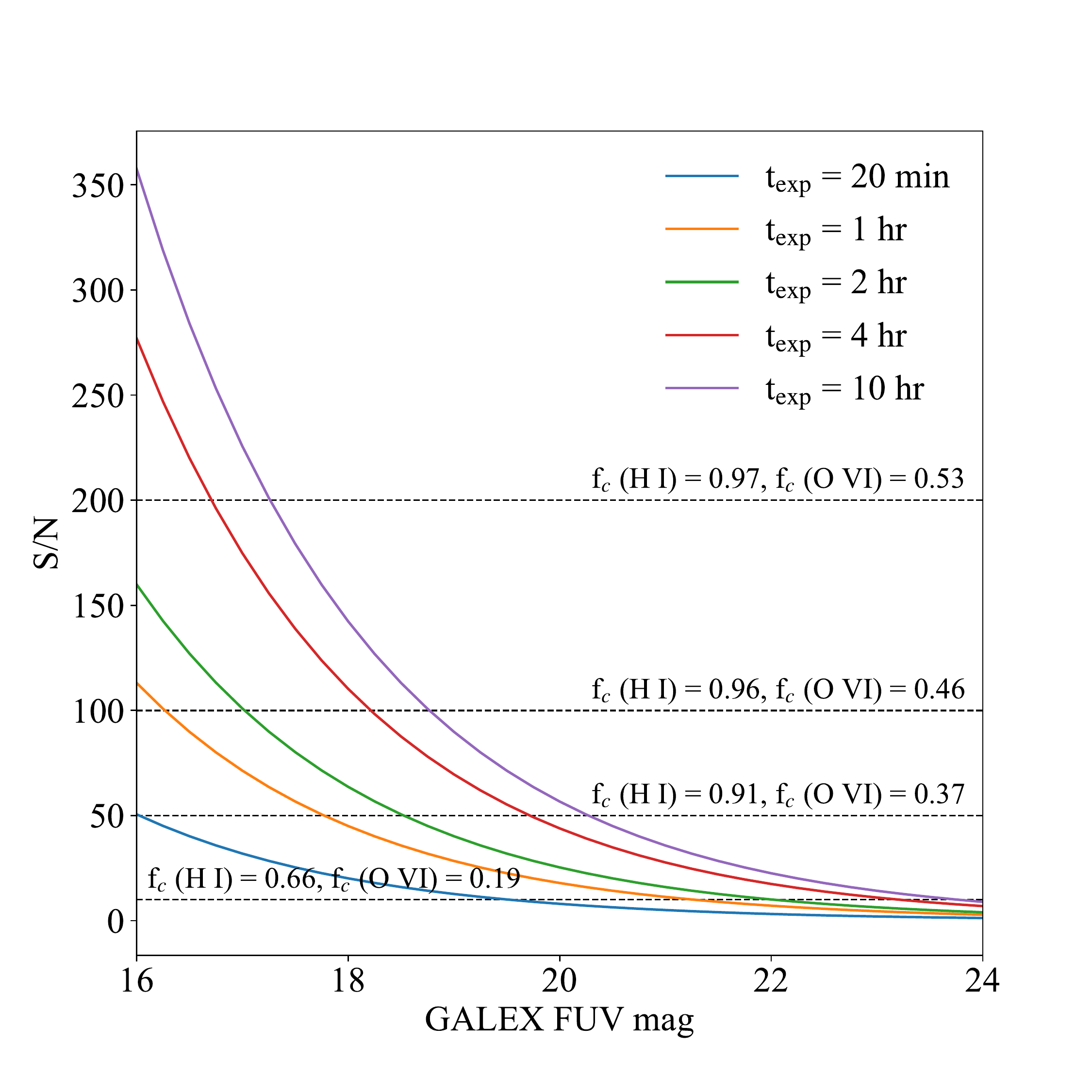}
    \caption{Signal-to-noise (S/N) ratio achieved under the current \emph{LUVOIR} 15m telescope concept when observing UV-bright sources at various exposure times (color coding indicated in legend).  The horizontal lines indicate covering fractions of \hone\ and \osix\ assuming the column densities detectable of each ion at the indicated S/N (see Figure \ref{fig:colTempDetect}).    
    \label{fig:luvoirSn}}
\end{figure}

The S/N of a far-UV spectrum is a function of the brightness of the background source and the integration time. 
In Figure \ref{fig:luvoirSn}, we show the predicted S/N of LUVOIR as a function of the GALEX FUV magnitude for 
different exposure times. The horizontal dashed lines show the covering fractions of \hone\ and \osix\ 
in {\sc RomulusC} for several different S/N (assuming each S/N corresponds to the median detectable column densities in 
Figure \ref{fig:colTempDetect}). 

Currently, the most sensitive UV spectrograph to \hone\ and \osix\ at low and intermediate redshifts is the 
\emph{HST/COS}. It consistently reaches S/N of at least 10, and can reach S/N values as high as 75 (corresponding
to the full range of S/N lines in Figure \ref{fig:colTempDetect}). 
For example, as of the July 2018 data release of the Hubble Spectroscopic Legacy Archive \citep[HSLA;][]{Peeples:2017aa}, 
~300 sightlines have S/N $\geq 10$, ~100 sightlines have S/N$\geq 20$, and 16 sightlines have S/N$\geq 50$. 
Using the covering fraction estimates presented in Figure \ref{fig:luvoirSn}, we predict that \emph{HST/COS}
can detect minimum covering fractions of 0.66 for \hone\ and 0.19 for \osix.

The next-generation UV instruments are still in the concept phase, vying for recommendation by the
Astro2020 Decadal Survey. Current examples include The Large Ultraviolet Optical Infrared (\emph{LUVOIR}) observatory \citep{Bolcar:2017aa}, which would feature a 15 meter aperture, as well as a host of instruments spanning the UV to the infrared, 
such as the Large Ultraviolet Multi-object Spectrograph \citep[LUMOS;][]{Harris:2018aa}. Although facilities
of this magnitude would likely not launch until the 2030s, we explore their anticipated advancements of UV cluster observations below, with the goal of understanding the prospect of studying the multiphase ICM in the coming decades. 

\emph{LUVOIR} would dramatically increase the S/N and observable
column density limits. For example, achieving the nominal S/N$=10$ we have assumed throughout the work
requires 5 hours of exposure time with \emph{HST/COS} for a source with $m_{FUV} = 19.5$. 
Achieving a similar S/N on \emph{LUVOIR} would only take 20 minutes. Furthermore, for modest exposure times
of 2-4 hours, \emph{LUVOIR} can deliver S/N$\ge 50$ (corresponding to covering fractions of 0.91 for \hone\ 
and 0.37 for \osix) for a large fraction of sources. However, according to our predictions, increasing the S/N to $\geq 100$ brings marginal returns. 

The transformational leap beyond current capabilities will likely come from the ability to observe
a larger sample of background sources. At fainter magnitudes ($m_{FUV} > 20$), the density of background 
sources (e.g., QSOs and star-forming galaxies) in the sky increases sharply \citep{Rubin:2018aa}. 
For example by cross-matching the UV Ultra Deep Field catalog \citep{Rafelski:2015aa} with the GALEX UV
source catalog \citep{Bianchi:2017aa} at $z = 0.2 - 1$, we found that there are $\ge 5900$ sources
per square degree with $m_{FUV} \ge 22$. Considering clustocentric impact parameters within 2 Mpc, this
source density translates to $\ge 200$ possible sources \textit{per cluster}.  Therefore, one could easily 
subselect the cluster sample based on the properties of the cluster or study a particular cluster in exquisite 
detail, following a similar approach to that of \citet{Yoon:2012yu} and \citet{Yoon:2017aa} for Virgo and Coma,
respectively, but \textit{for virtually any clusters in the sky}.

\section{Conclusions}
\label{sec:conclusions}

We use the high-resolution hydrodynamical galaxy cluster simulation, {\sc RomulusC}, to examine the multiphase
structure of the intracluster medium (ICM) and its effect on the circumgalactic media (CGM) of constituent galaxies.
We focus our analysis on cool-warm ($T = 10^{4-6}$K) gas at the cluster outskirts which is particularly
difficult to observe in X-rays. We show that this gas can be observed in the ultraviolet (UV) with \emph{HST/COS} and make
predictions for observations with next-generation telescopes, like \emph{LUVOIR}. We summarize our key findings as follows:
\begin{enumerate}
\item At all clustocentric radii, we find significant quantities of multiphase gas in the cool ($10^{4-5}$K) and warm ($10^{5-6}$K) gas phase, observable in the UV. Relative to the hot ($\mathrm{T}>10^6$K), X-ray emitting gas, the share of cool-warm material increases rapidly toward the cluster outskirts.  By $r > 2.5 \mathrm{R}_{200}$, the warm gas mass fraction exceeds that of the hot gas, and even the cool gas exceeds the hot gas at $r > 3.5 \mathrm{R}_{200}$.
\item The ICM is marked by a wide range of metallicity from $Z\approx 0.01 Z_{\odot}$ to $Z> Z_{\odot}$.  Our simulations predict that the gas at temperatures and densities probed by X-rays exhibits metallicities broadly consistent with X-ray measurements at $r < \mathrm{R}_{200}$; however, the lower temperature and density regime to which the UV is sensitive possesses a much wider range in metallicity, including lower metallicities from the well-mixed ICM and highly enriched gas that is stripped from the CGM of infalling galaxies.
\item We predict detection rates of \hone, \cfour, and \osix\, which sample different temperatures of the cool-warm gas,
throughout the ICM. We find that the detection profiles of all three ions decline from the cluster center out to 1 Mpc, but
remain relatively flat out to the edge of the simulated volume, at 3 Mpc.

\item The CGM of cluster galaxies are readily stripped during the early stages of infall. CGM stripping is apparent in
galaxies as far as $4\mathrm{R}_{200}$ from the cluster center and becomes progressively more pronounced near the cluster center.  
This stripping is observable in the the halo-centric covering fractions of \hone, \cfour, and \osix\ for galaxies at various distances from the cluster center.

\item To test the feasibility of confronting our galaxy cluster simulation and the physics therein with observations, we generate and analyze synthetic UV spectra of sightlines piercing the {\sc RomulusC} ICM and CGM of cluster galaxies.  Within 16 synthetic sightlines, we detect 19 components of \hone, \osix, and \cthree, including several broad \lya\ components tracing the elusive warm gas regime, in various environments from the cluster center to the outskirts.  These synthetic data demonstrate that a large UV spectroscopic survey using currently available instrumentation can effectively be leveraged to test cluster formation and evolution models.
\item Considering current and future UV spectroscopic capabilities, we examine potential observing strategies to constrain the multiphase ICM and cluster galaxy CGM.  Substantive, important progress can be made now with \emph{HST/COS} given modest investments of telescope time, for an ensemble of clusters probed by UV-bright ($m_{\rm FUV}<19$) QSOs to constrain the cool-warm gas profiles, metal enrichment, and CGM in cluster models.  Dramatically improved diagnostic power would be afforded by a large-aperture space-borne UV-sensitive telescope such as the \emph{LUVOIR} concept currently being considered.  For the same ensemble of clusters one could observe at S/N$\sim10$ with \emph{HST/COS}, \emph{LUVOIR} would deliver S/N$=50-100$ and detect extremely diffuse cool-warm gas approaching detection rates of $90\%$.
\end{enumerate}

UV spectroscopy provides unique insights into the cold-warm gas in and around the most massive structures in the Universe and offers a highly complementary view of the baryonic contents of galaxy clusters derived from X-ray and microwave observations. Together, these forthcoming multiwavelength observations will provide a comprehensive view of the multiphase gas in the ICM of galaxy clusters and the CGM of cluster galaxies to transform our understanding of galaxy cluster physics and galaxy evolution.

\section*{Acknowledgements}

IB would like to thank Cameron Hummels, Nicole Sanchez, and Akaxia Cruz for their insightful conversations regarding this work. 
This research is part of the Blue Waters sustained-petascale computing project, which is supported by the National Science Foundation (awards OCI-0725070 and ACI-1238993) and the state of Illinois. Blue Waters is a joint effort of the University of Illinois at Urbana-Champaign and its National Center for Supercomputing Applications. This work is also part of a Petascale Computing Resource Allocations allocation support by the National Science Foundation (award number OAC-1613674).  This work also used the Extreme Science and Engineering Discovery Environment (XSEDE), which is supported by National Science Foundation grant number ACI-1548562. Resources supporting this work were also provided by the NASA High-End Computing (HEC) Program through the NASA Advanced Supercomputing (NAS) Division at Ames Research Center.

ISB,  TQ  and  MT  were  partially  supported  by NSF award AST-1514868.
ISB  received  partial  support  from NASA award HST-AR-15046. MT gratefully acknowledges support from the YCAA Prize Postdoctoral Fellowship.


\bibliographystyle{mnras}

\bibliography{ButskyReferences}

\begin{thebibliography}{}
\makeatletter
\relax
\def\mn@urlcharsother{\let\do\@makeother \do\$\do\&\do\#\do\^\do\_\do\%\do\~}
\def\mn@doi{\begingroup\mn@urlcharsother \@ifnextchar [ {\mn@doi@}
  {\mn@doi@[]}}
\def\mn@doi@[#1]#2{\def\@tempa{#1}\ifx\@tempa\@empty \href
  {http://dx.doi.org/#2} {doi:#2}\else \href {http://dx.doi.org/#2} {#1}\fi
  \endgroup}
\def\mn@eprint#1#2{\mn@eprint@#1:#2::\@nil}
\def\mn@eprint@arXiv#1{\href {http://arxiv.org/abs/#1} {{\tt arXiv:#1}}}
\def\mn@eprint@dblp#1{\href {http://dblp.uni-trier.de/rec/bibtex/#1.xml}
  {dblp:#1}}
\def\mn@eprint@#1:#2:#3:#4\@nil{\def\@tempa {#1}\def\@tempb {#2}\def\@tempc
  {#3}\ifx \@tempc \@empty \let \@tempc \@tempb \let \@tempb \@tempa \fi \ifx
  \@tempb \@empty \def\@tempb {arXiv}\fi \@ifundefined
  {mn@eprint@\@tempb}{\@tempb:\@tempc}{\expandafter \expandafter \csname
  mn@eprint@\@tempb\endcsname \expandafter{\@tempc}}}

\bibitem[\protect\citeauthoryear{{Allen}, {Evrard}  \& {Mantz}}{{Allen}
  et~al.}{2011}]{Allen:2011aa}
{Allen} S.~W.,  {Evrard} A.~E.,   {Mantz} A.~B.,  2011, \araa, 49, 409

\bibitem[\protect\citeauthoryear{{Asplund}, {Grevesse}, {Sauval}  \&
  {Scott}}{{Asplund} et~al.}{2009}]{Asplund:2009}
{Asplund} M.,  {Grevesse} N.,  {Sauval} A.~J.,   {Scott} P.,  2009, \mn@doi
  [\araa] {10.1146/annurev.astro.46.060407.145222}, \href
  {http://adsabs.harvard.edu/abs/2009ARA%26A..47..481A} {47, 481}

\bibitem[\protect\citeauthoryear{{Bah{\'e}} et~al.,}{{Bah{\'e}}
  et~al.}{2017}]{Bahe:2017aa}
{Bah{\'e}} Y.~M.,  et~al., 2017, \mnras, 470, 4186

\bibitem[\protect\citeauthoryear{{Bianchi}, {Shiao}  \& {Thilker}}{{Bianchi}
  et~al.}{2017}]{Bianchi:2017aa}
{Bianchi} L.,  {Shiao} B.,   {Thilker} D.,  2017, The Astrophysical Journal
  Supplement Series, 230, 24

\bibitem[\protect\citeauthoryear{{Bolcar} et~al.,}{{Bolcar}
  et~al.}{2017}]{Bolcar:2017aa}
{Bolcar} M.~R.,  et~al., 2017, in Society of Photo-Optical Instrumentation
  Engineers (SPIE) Conference Series. p. 1039809

\bibitem[\protect\citeauthoryear{{Booth} \& {Schaye}}{{Booth} \&
  {Schaye}}{2009}]{Booth:2009}
{Booth} C.~M.,  {Schaye} J.,  2009, \mn@doi [\mnras]
  {10.1111/j.1365-2966.2009.15043.x}, \href
  {http://adsabs.harvard.edu/abs/2009MNRAS.398...53B} {398, 53}

\bibitem[\protect\citeauthoryear{{Bordoloi} et~al.,}{{Bordoloi}
  et~al.}{2014}]{Bordoloi:2014}
{Bordoloi} R.,  et~al., 2014, \mn@doi [\apj] {10.1088/0004-637X/796/2/136},
  \href {http://adsabs.harvard.edu/abs/2014ApJ...796..136B} {796, 136}

\bibitem[\protect\citeauthoryear{Burchett et~al.,}{Burchett
  et~al.}{2015}]{Burchett:2015rf}
Burchett J.~N.,  et~al., 2015, ApJ, 815, 91

\bibitem[\protect\citeauthoryear{{Burchett}, {Tripp}, {Wang}, {Willmer},
  {Bowen}  \& {Jenkins}}{{Burchett} et~al.}{2018}]{Burchett:2018aa}
{Burchett} J.~N.,  {Tripp} T.~M.,  {Wang} Q.~D.,  {Willmer} C.~N.~A.,  {Bowen}
  D.~V.,   {Jenkins} E.~B.,  2018, MNRAS, 475, 2067

\bibitem[\protect\citeauthoryear{Cen \& Ostriker}{Cen \&
  Ostriker}{1999}]{Cen:1999yq}
Cen R.,  Ostriker J.~P.,  1999, ApJ, 514, 1

\bibitem[\protect\citeauthoryear{{Davies} \& {Lewis}}{{Davies} \&
  {Lewis}}{1973}]{Davies:1973}
{Davies} R.~D.,  {Lewis} B.~M.,  1973, \mn@doi [\mnras]
  {10.1093/mnras/165.2.231}, \href
  {http://adsabs.harvard.edu/abs/1973MNRAS.165..231D} {165, 231}

\bibitem[\protect\citeauthoryear{{Dressler}}{{Dressler}}{1980}]{Dressler:1980}
{Dressler} A.,  1980, \mn@doi [\apj] {10.1086/157753}, \href
  {http://adsabs.harvard.edu/abs/1980ApJ...236..351D} {236, 351}

\bibitem[\protect\citeauthoryear{{Emerick}, {Bryan}  \& {Putman}}{{Emerick}
  et~al.}{2015}]{Emerick:2015}
{Emerick} A.,  {Bryan} G.,   {Putman} M.~E.,  2015, \mn@doi [\mnras]
  {10.1093/mnras/stv1936}, \href
  {http://adsabs.harvard.edu/abs/2015MNRAS.453.4051E} {453, 4051}

\bibitem[\protect\citeauthoryear{{Emerick}, {Bryan}  \& {Mac Low}}{{Emerick}
  et~al.}{2019}]{Emerick:2019}
{Emerick} A.,  {Bryan} G.~L.,   {Mac Low} M.-M.,  2019, \mn@doi [\mnras]
  {10.1093/mnras/sty2689}, \href
  {http://adsabs.harvard.edu/abs/2019MNRAS.482.1304E} {482, 1304}

\bibitem[\protect\citeauthoryear{{Ferland} et~al.,}{{Ferland}
  et~al.}{2013}]{Ferland:2013}
{Ferland} G.~J.,  et~al., 2013, Rev. Mexicana Astron. Astrofis., \href
  {http://adsabs.harvard.edu/abs/2013RMxAA..49..137F} {49, 137}

\bibitem[\protect\citeauthoryear{{Fraser-McKelvie}, {Brown}, {Pimbblet},
  {Dolley}  \& {Bonne}}{{Fraser-McKelvie} et~al.}{2018}]{Fraser:2017}
{Fraser-McKelvie} A.,  {Brown} M.~J.~I.,  {Pimbblet} K.,  {Dolley} T.,
  {Bonne} N.~J.,  2018, \mn@doi [\mnras] {10.1093/mnras/stx2823}, \href
  {http://adsabs.harvard.edu/abs/2018MNRAS.474.1909F} {474, 1909}

\bibitem[\protect\citeauthoryear{{Fumagalli}, {Fossati}, {Hau}, {Gavazzi},
  {Bower}, {Sun}  \& {Boselli}}{{Fumagalli} et~al.}{2014}]{Fumagalli:2014}
{Fumagalli} M.,  {Fossati} M.,  {Hau} G.~K.~T.,  {Gavazzi} G.,  {Bower} R.,
  {Sun} M.,   {Boselli} A.,  2014, \mn@doi [\mnras] {10.1093/mnras/stu2092},
  \href {http://adsabs.harvard.edu/abs/2014MNRAS.445.4335F} {445, 4335}

\bibitem[\protect\citeauthoryear{Ge, Wang, Tripp, Li, Gu  \& Ji}{Ge
  et~al.}{2016}]{Ge:2016lr}
Ge C.,  Wang Q.~D.,  Tripp T.~M.,  Li Z.,  Gu Q.,   Ji L.,  2016, MNRAS, 459,
  366

\bibitem[\protect\citeauthoryear{{Ghavamian} et~al.,}{{Ghavamian}
  et~al.}{2009}]{COSLSF}
{Ghavamian} P.,  et~al., 2009, Technical report, {Preliminary Characterization
  of the Post- Launch Line Spread Function of COS}

\bibitem[\protect\citeauthoryear{Gnat \& Sternberg}{Gnat \&
  Sternberg}{2007}]{Gnat:2007fk}
Gnat O.,  Sternberg A.,  2007, ApJS, 168, 213

\bibitem[\protect\citeauthoryear{{Gunn} \& {Gott}}{{Gunn} \&
  {Gott}}{1972}]{Gunn:1972}
{Gunn} J.~E.,  {Gott} III J.~R.,  1972, \mn@doi [\apj] {10.1086/151605}, \href
  {http://adsabs.harvard.edu/abs/1972ApJ...176....1G} {176, 1}

\bibitem[\protect\citeauthoryear{{Haardt} \& {Madau}}{{Haardt} \&
  {Madau}}{2012}]{HaardtMadau:2012}
{Haardt} F.,  {Madau} P.,  2012, \mn@doi [\apj] {10.1088/0004-637X/746/2/125},
  \href {https://ui.adsabs.harvard.edu/\#abs/2012ApJ...746..125H} {746, 125}

\bibitem[\protect\citeauthoryear{{Harris}, {France}, {Fleming}  \&
  {Bolcar}}{{Harris} et~al.}{2018}]{Harris:2018aa}
{Harris} W.,  {France} K.,  {Fleming} B.,   {Bolcar} M.,  2018, AGU Fall
  Meeting Abstracts

\bibitem[\protect\citeauthoryear{{Hopkins} et~al.,}{{Hopkins}
  et~al.}{2018}]{Hopkins:2018}
{Hopkins} P.~F.,  et~al., 2018, \mn@doi [\mnras] {10.1093/mnras/sty1690}, \href
  {http://adsabs.harvard.edu/abs/2018MNRAS.480..800H} {480, 800}

\bibitem[\protect\citeauthoryear{{Hummels}, {Smith}  \& {Silvia}}{{Hummels}
  et~al.}{2017}]{Hummels:2017}
{Hummels} C.~B.,  {Smith} B.~D.,   {Silvia} D.~W.,  2017, \mn@doi [\apj]
  {10.3847/1538-4357/aa7e2d}, \href
  {http://adsabs.harvard.edu/abs/2017ApJ...847...59H} {847, 59}

\bibitem[\protect\citeauthoryear{{Hummels} et~al.,}{{Hummels}
  et~al.}{2018}]{Hummels:2018}
{Hummels} C.~B.,  et~al., 2018, arXiv e-prints, \href
  {http://adsabs.harvard.edu/abs/2018arXiv181112410H} {}

\bibitem[\protect\citeauthoryear{{J{\'a}chym}, {Combes}, {Cortese}, {Sun}  \&
  {Kenney}}{{J{\'a}chym} et~al.}{2014}]{Jachym:2014}
{J{\'a}chym} P.,  {Combes} F.,  {Cortese} L.,  {Sun} M.,   {Kenney} J.~D.~P.,
  2014, \mn@doi [\apj] {10.1088/0004-637X/792/1/11}, \href
  {http://adsabs.harvard.edu/abs/2014ApJ...792...11J} {792, 11}

\bibitem[\protect\citeauthoryear{{Johnson}, {Chen}  \& {Mulchaey}}{{Johnson}
  et~al.}{2015}]{Johnson:2015}
{Johnson} S.~D.,  {Chen} H.-W.,   {Mulchaey} J.~S.,  2015, \mn@doi [\mnras]
  {10.1093/mnras/stv553}, \href
  {http://adsabs.harvard.edu/abs/2015MNRAS.449.3263J} {449, 3263}

\bibitem[\protect\citeauthoryear{{Kravtsov} \& {Borgani}}{{Kravtsov} \&
  {Borgani}}{2012}]{Kravtsov:2012}
{Kravtsov} A.~V.,  {Borgani} S.,  2012, \mn@doi [\araa]
  {10.1146/annurev-astro-081811-125502}, \href
  {http://adsabs.harvard.edu/abs/2012ARA%26A..50..353K} {50, 353}

\bibitem[\protect\citeauthoryear{{Lake}, {Katz}  \& {Moore}}{{Lake}
  et~al.}{1998}]{Lake:1998}
{Lake} G.,  {Katz} N.,   {Moore} B.,  1998, \mn@doi [\apj] {10.1086/305265},
  \href {http://adsabs.harvard.edu/abs/1998ApJ...495..152L} {495, 152}

\bibitem[\protect\citeauthoryear{Lau, Kravtsov  \& Nagai}{Lau
  et~al.}{2009}]{Lau:2009qy}
Lau E.~T.,  Kravtsov A.~V.,   Nagai D.,  2009, ApJ, 705, 1129

\bibitem[\protect\citeauthoryear{{Lau}, {Nagai}, {Avestruz}, {Nelson}  \&
  {Vikhlinin}}{{Lau} et~al.}{2015}]{Lau2015}
{Lau} E.~T.,  {Nagai} D.,  {Avestruz} C.,  {Nelson} K.,   {Vikhlinin} A.,
  2015, \mn@doi [\apj] {10.1088/0004-637X/806/1/68}, \href
  {http://adsabs.harvard.edu/abs/2015ApJ...806...68L} {806, 68}

\bibitem[\protect\citeauthoryear{Lehner, Savage, Richter, Sembach, Tripp  \&
  Wakker}{Lehner et~al.}{2007}]{Lehner:2007lr}
Lehner N.,  Savage B.~D.,  Richter P.,  Sembach K.~R.,  Tripp T.~M.,   Wakker
  B.~P.,  2007, ApJ, 658, 680

\bibitem[\protect\citeauthoryear{{Menon}, {Wesolowski}, {Zheng}, {Jetley},
  {Kale}, {Quinn}  \& {Governato}}{{Menon} et~al.}{2015}]{Menon:2015}
{Menon} L.,  {Wesolowski} F.,  {Zheng} G.,  {Jetley} P.,  {Kale} L.~V.,
  {Quinn} T.~R.,   {Governato} F.,  2015, in Adaptive Techniques for Clustered
  N-Body Cosmological Simulations. Computational Astrophysics and Cosmology, 2,
  1..

\bibitem[\protect\citeauthoryear{{Mernier} et~al.,}{{Mernier}
  et~al.}{2018}]{Mernier:2018ab}
{Mernier} F.,  et~al., 2018, \ssr, 214, 129

\bibitem[\protect\citeauthoryear{{Moore}, {Katz}, {Lake}, {Dressler}  \&
  {Oemler}}{{Moore} et~al.}{1996}]{Moore:1996}
{Moore} B.,  {Katz} N.,  {Lake} G.,  {Dressler} A.,   {Oemler} A.,  1996,
  \mn@doi [\nat] {10.1038/379613a0}, \href
  {http://adsabs.harvard.edu/abs/1996Natur.379..613M} {379, 613}

\bibitem[\protect\citeauthoryear{{Mroczkowski} et~al.,}{{Mroczkowski}
  et~al.}{2019}]{Mroczkowski:2019aa}
{Mroczkowski} T.,  et~al., 2019, \ssr, 215, 17

\bibitem[\protect\citeauthoryear{Muzahid, Charlton, Nagai, Schaye  \&
  Srianand}{Muzahid et~al.}{2017}]{Muzahid:2017lr}
Muzahid S.,  Charlton J.,  Nagai D.,  Schaye J.,   Srianand R.,  2017, The
  Astrophysical Journal Letters, 846, L8

\bibitem[\protect\citeauthoryear{{Nagai} \& {Lau}}{{Nagai} \&
  {Lau}}{2011}]{Nagai:2011}
{Nagai} D.,  {Lau} E.~T.,  2011, \mn@doi [\apjl] {10.1088/2041-8205/731/1/L10},
  \href {http://adsabs.harvard.edu/abs/2011ApJ...731L..10N} {731, L10}

\bibitem[\protect\citeauthoryear{{Nagai}, {Lau}, {Avestruz}, {Nelson}  \&
  {Rudd}}{{Nagai} et~al.}{2013}]{Nagai:2013aa}
{Nagai} D.,  {Lau} E.~T.,  {Avestruz} C.,  {Nelson} K.,   {Rudd} D.~H.,  2013,
  \apj, 777, 137

\bibitem[\protect\citeauthoryear{Nelson, Lau  \& Nagai}{Nelson
  et~al.}{2014}]{Nelson:2014aa}
Nelson K.,  Lau E.~T.,   Nagai D.,  2014, ApJ, 792, 25

\bibitem[\protect\citeauthoryear{{Nelson} et~al.,}{{Nelson}
  et~al.}{2019}]{Nelson:2019}
{Nelson} D.,  et~al., 2019, arXiv e-prints, \href
  {http://adsabs.harvard.edu/abs/2019arXiv190205554N} {}

\bibitem[\protect\citeauthoryear{{Nielsen}, {Kacprzak}, {Pointon}, {Churchill}
  \& {Murphy}}{{Nielsen} et~al.}{2018}]{Nielsen:2018}
{Nielsen} N.~M.,  {Kacprzak} G.~G.,  {Pointon} S.~K.,  {Churchill} C.~W.,
  {Murphy} M.~T.,  2018, \mn@doi [\apj] {10.3847/1538-4357/aaedbd}, \href
  {http://adsabs.harvard.edu/abs/2018ApJ...869..153N} {869, 153}

\bibitem[\protect\citeauthoryear{{Oppenheimer} \& {Schaye}}{{Oppenheimer} \&
  {Schaye}}{2013}]{Oppenheimer:2013}
{Oppenheimer} B.~D.,  {Schaye} J.,  2013, \mn@doi [\mnras]
  {10.1093/mnras/stt1043}, \href
  {http://adsabs.harvard.edu/abs/2013MNRAS.434.1043O} {434, 1043}

\bibitem[\protect\citeauthoryear{{Oppenheimer}, {Segers}, {Schaye}, {Richings}
  \& {Crain}}{{Oppenheimer} et~al.}{2018}]{Oppenheimer:2018}
{Oppenheimer} B.~D.,  {Segers} M.,  {Schaye} J.,  {Richings} A.~J.,   {Crain}
  R.~A.,  2018, \mn@doi [\mnras] {10.1093/mnras/stx2967}, \href
  {http://adsabs.harvard.edu/abs/2018MNRAS.474.4740O} {474, 4740}

\bibitem[\protect\citeauthoryear{{Peeples} et~al.,}{{Peeples}
  et~al.}{2017}]{Peeples:2017aa}
{Peeples} M.,  et~al., 2017, Technical report, {The Hubble Spectroscopic Legacy
  Archive}

\bibitem[\protect\citeauthoryear{{Planck Collaboration} et~al.,}{{Planck
  Collaboration} et~al.}{2016}]{Planck-Collaboration:2016eq}
{Planck Collaboration} et~al., 2016, Astronomy and Astrophysics, 594, A13

\bibitem[\protect\citeauthoryear{{Pointon}, {Nielsen}, {Kacprzak}, {Muzahid},
  {Churchill}  \& {Charlton}}{{Pointon} et~al.}{2017}]{Pointon:2017}
{Pointon} S.~K.,  {Nielsen} N.~M.,  {Kacprzak} G.~G.,  {Muzahid} S.,
  {Churchill} C.~W.,   {Charlton} J.~C.,  2017, \mn@doi [\apj]
  {10.3847/1538-4357/aa7743}, \href
  {http://adsabs.harvard.edu/abs/2017ApJ...844...23P} {844, 23}

\bibitem[\protect\citeauthoryear{{Pontzen} \& {Tremmel}}{{Pontzen} \&
  {Tremmel}}{2018}]{Pontzen:2018}
{Pontzen} A.,  {Tremmel} M.,  2018, \mn@doi [\apjs] {10.3847/1538-4365/aac832},
  \href {http://adsabs.harvard.edu/abs/2018ApJS..237...23P} {237, 23}

\bibitem[\protect\citeauthoryear{{Pontzen}, {Ro{\v s}kar}, {Stinson}, {Woods},
  {Reed}, {Coles}  \& {Quinn}}{{Pontzen} et~al.}{2013}]{pynbody}
{Pontzen} A.,  {Ro{\v s}kar} R.,  {Stinson} G.~S.,  {Woods} R.,  {Reed} D.~M.,
  {Coles} J.,   {Quinn} T.~R.,  2013, {pynbody: Astrophysics Simulation
  Analysis for Python}

\bibitem[\protect\citeauthoryear{{Pontzen}, {Tremmel}, {Roth}, {Peiris},
  {Saintonge}, {Volonteri}, {Quinn}  \& {Governato}}{{Pontzen}
  et~al.}{2017}]{Pontzen:2017}
{Pontzen} A.,  {Tremmel} M.,  {Roth} N.,  {Peiris} H.~V.,  {Saintonge} A.,
  {Volonteri} M.,  {Quinn} T.,   {Governato} F.,  2017, \mn@doi [\mnras]
  {10.1093/mnras/stw2627}, \href
  {http://adsabs.harvard.edu/abs/2017MNRAS.465..547P} {465, 547}

\bibitem[\protect\citeauthoryear{{Pratt}, {Arnaud}, {Biviano}, {Eckert},
  {Ettori}, {Nagai}, {Okabe}  \& {Reiprich}}{{Pratt}
  et~al.}{2019}]{Pratt:2019aa}
{Pratt} G.~W.,  {Arnaud} M.,  {Biviano} A.,  {Eckert} D.,  {Ettori} S.,
  {Nagai} D.,  {Okabe} N.,   {Reiprich} T.~H.,  2019, \ssr, 215, 25

\bibitem[\protect\citeauthoryear{{Rafelski} et~al.,}{{Rafelski}
  et~al.}{2015}]{Rafelski:2015aa}
{Rafelski} M.,  et~al., 2015, \aj, 150, 31

\bibitem[\protect\citeauthoryear{{Rasia} et~al.,}{{Rasia}
  et~al.}{2014}]{Rasia2014}
{Rasia} E.,  et~al., 2014, \mn@doi [\apj] {10.1088/0004-637X/791/2/96}, \href
  {http://adsabs.harvard.edu/abs/2014ApJ...791...96R} {791, 96}

\bibitem[\protect\citeauthoryear{{Roediger} et~al.,}{{Roediger}
  et~al.}{2015}]{Roediger:2015}
{Roediger} E.,  et~al., 2015, \mn@doi [\apj] {10.1088/0004-637X/806/1/104},
  \href {https://ui.adsabs.harvard.edu/\#abs/2015ApJ...806..104R} {806, 104}

\bibitem[\protect\citeauthoryear{{Rubin}, {Diamond-Stanic}, {Coil}, {Crighton}
  \& {Moustakas}}{{Rubin} et~al.}{2018}]{Rubin:2018aa}
{Rubin} K.~H.~R.,  {Diamond-Stanic} A.~M.,  {Coil} A.~L.,  {Crighton} N.~H.~M.,
    {Moustakas} J.,  2018, ApJ, 853, 95

\bibitem[\protect\citeauthoryear{{Sanchez}, {Werk}, {Tremmel}, {Pontzen},
  {Christensen}, {Quinn}  \& {Cruz}}{{Sanchez} et~al.}{2018}]{Sanchez:2018}
{Sanchez} N.~N.,  {Werk} J.~K.,  {Tremmel} M.,  {Pontzen} A.,  {Christensen}
  C.,  {Quinn} T.,   {Cruz} A.,  2018, arXiv e-prints, \href
  {http://adsabs.harvard.edu/abs/2018arXiv181012319S} {}

\bibitem[\protect\citeauthoryear{{Sandage}, {Binggeli}  \& {Tammann}}{{Sandage}
  et~al.}{1985}]{Sandage:1985}
{Sandage} A.,  {Binggeli} B.,   {Tammann} G.~A.,  1985, \mn@doi [\aj]
  {10.1086/113875}, \href {http://adsabs.harvard.edu/abs/1985AJ.....90.1759S}
  {90, 1759}

\bibitem[\protect\citeauthoryear{{Shen}, {Wadsley}  \& {Stinson}}{{Shen}
  et~al.}{2010}]{Shen:2010}
{Shen} S.,  {Wadsley} J.,   {Stinson} G.,  2010, \mn@doi [\mnras]
  {10.1111/j.1365-2966.2010.17047.x}, \href
  {http://adsabs.harvard.edu/abs/2010MNRAS.407.1581S} {407, 1581}

\bibitem[\protect\citeauthoryear{{Stinson}, {Seth}, {Katz}, {Wadsley},
  {Governato}  \& {Quinn}}{{Stinson} et~al.}{2006}]{Stinson:2006}
{Stinson} G.,  {Seth} A.,  {Katz} N.,  {Wadsley} J.,  {Governato} F.,   {Quinn}
  T.,  2006, \mn@doi [\mnras] {10.1111/j.1365-2966.2006.11097.x}, \href
  {http://adsabs.harvard.edu/abs/2006MNRAS.373.1074S} {373, 1074}

\bibitem[\protect\citeauthoryear{{Stocke}, {Keeney}, {Danforth}, {Shull},
  {Froning}, {Green}, {Penton}  \& {Savage}}{{Stocke}
  et~al.}{2013}]{Stocke:2013}
{Stocke} J.~T.,  {Keeney} B.~A.,  {Danforth} C.~W.,  {Shull} J.~M.,  {Froning}
  C.~S.,  {Green} J.~C.,  {Penton} S.~V.,   {Savage} B.~D.,  2013, \mn@doi
  [\apj] {10.1088/0004-637X/763/2/148}, \href
  {http://adsabs.harvard.edu/abs/2013ApJ...763..148S} {763, 148}

\bibitem[\protect\citeauthoryear{{Stocke}, {Keeney}, {Danforth}, {Oppenheimer},
  {Pratt}, {Berlind}, {Impey}  \& {Jannuzi}}{{Stocke}
  et~al.}{2019}]{Stocke:2019}
{Stocke} J.~T.,  {Keeney} B.~A.,  {Danforth} C.~W.,  {Oppenheimer} B.~D.,
  {Pratt} C.~T.,  {Berlind} A.~A.,  {Impey} C.,   {Jannuzi} B.,  2019, \mn@doi
  [\apjs] {10.3847/1538-4365/aaf73d}, \href
  {http://adsabs.harvard.edu/abs/2019ApJS..240...15S} {240, 15}

\bibitem[\protect\citeauthoryear{Tejos et~al.,}{Tejos
  et~al.}{2016}]{Tejos:2016qv}
Tejos N.,  et~al., 2016, MNRAS, 455, 2662

\bibitem[\protect\citeauthoryear{Tepper-Garc{\'\i}a, Richter, Schaye, Booth,
  Dalla~Vecchia  \& Theuns}{Tepper-Garc{\'\i}a
  et~al.}{2012}]{Tepper-Garcia:2012aa}
Tepper-Garc{\'\i}a T.,  Richter P.,  Schaye J.,  Booth C.~M.,  Dalla~Vecchia
  C.,   Theuns T.,  2012, MNRAS, 425, 1640

\bibitem[\protect\citeauthoryear{{Tonnesen}, {Bryan}  \& {van
  Gorkom}}{{Tonnesen} et~al.}{2007}]{Tonnesen:2007}
{Tonnesen} S.,  {Bryan} G.~L.,   {van Gorkom} J.~H.,  2007, \mn@doi [\apj]
  {10.1086/523034}, \href {http://adsabs.harvard.edu/abs/2007ApJ...671.1434T}
  {671, 1434}

\bibitem[\protect\citeauthoryear{{Toomre} \& {Toomre}}{{Toomre} \&
  {Toomre}}{1972}]{Toomre:1972}
{Toomre} A.,  {Toomre} J.,  1972, \mn@doi [\apj] {10.1086/151823}, \href
  {http://adsabs.harvard.edu/abs/1972ApJ...178..623T} {178, 623}

\bibitem[\protect\citeauthoryear{{Tremmel}, {Governato}, {Volonteri}  \&
  {Quinn}}{{Tremmel} et~al.}{2015}]{Tremmel:2015}
{Tremmel} M.,  {Governato} F.,  {Volonteri} M.,   {Quinn} T.~R.,  2015, \mn@doi
  [\mnras] {10.1093/mnras/stv1060}, \href
  {http://adsabs.harvard.edu/abs/2015MNRAS.451.1868T} {451, 1868}

\bibitem[\protect\citeauthoryear{{Tremmel}, {Karcher}, {Governato},
  {Volonteri}, {Quinn}, {Pontzen}, {Anderson}  \& {Bellovary}}{{Tremmel}
  et~al.}{2017}]{Tremmel:2017}
{Tremmel} M.,  {Karcher} M.,  {Governato} F.,  {Volonteri} M.,  {Quinn} T.~R.,
  {Pontzen} A.,  {Anderson} L.,   {Bellovary} J.,  2017, \mn@doi [\mnras]
  {10.1093/mnras/stx1160}, \href
  {http://adsabs.harvard.edu/abs/2017MNRAS.470.1121T} {470, 1121}

\bibitem[\protect\citeauthoryear{{Tremmel} et~al.,}{{Tremmel}
  et~al.}{2019}]{Tremmel:2019}
{Tremmel} M.,  et~al., 2019, \mn@doi [\mnras] {10.1093/mnras/sty3336}, \href
  {http://adsabs.harvard.edu/abs/2019MNRAS.483.3336T} {483, 3336}

\bibitem[\protect\citeauthoryear{{Tripp} et~al.,}{{Tripp}
  et~al.}{2011}]{tripp11}
{Tripp} T.~M.,  et~al., 2011, \mn@doi [Science] {10.1126/science.1209850},
  \href {http://adsabs.harvard.edu/abs/2011Sci...334..952T} {334, 952}

\bibitem[\protect\citeauthoryear{{Tumlinson} et~al.,}{{Tumlinson}
  et~al.}{2011}]{Tumlinson:2011}
{Tumlinson} J.,  et~al., 2011, \mn@doi [\apj] {10.1088/0004-637X/733/2/111},
  \href {http://adsabs.harvard.edu/abs/2011ApJ...733..111T} {733, 111}

\bibitem[\protect\citeauthoryear{{Tumlinson} et~al.,}{{Tumlinson}
  et~al.}{2013}]{Tumlinson:2013}
{Tumlinson} J.,  et~al., 2013, \mn@doi [\apj] {10.1088/0004-637X/777/1/59},
  \href {http://adsabs.harvard.edu/abs/2013ApJ...777...59T} {777, 59}

\bibitem[\protect\citeauthoryear{{Tumlinson}, {Peeples}  \& {Werk}}{{Tumlinson}
  et~al.}{2017}]{tumlinson17}
{Tumlinson} J.,  {Peeples} M.~S.,   {Werk} J.~K.,  2017, \mn@doi [\araa]
  {10.1146/annurev-astro-091916-055240}, \href
  {http://adsabs.harvard.edu/abs/2017ARA%26A..55..389T} {55, 389}

\bibitem[\protect\citeauthoryear{{Turk}, {Smith}, {Oishi}, {Skory}, {Skillman},
  {Abel}  \& {Norman}}{{Turk} et~al.}{2011}]{Turk:2011}
{Turk} M.~J.,  {Smith} B.~D.,  {Oishi} J.~S.,  {Skory} S.,  {Skillman} S.~W.,
  {Abel} T.,   {Norman} M.~L.,  2011, \mn@doi [\apjs]
  {10.1088/0067-0049/192/1/9}, \href
  {http://adsabs.harvard.edu/abs/2011ApJS..192....9T} {192, 9}

\bibitem[\protect\citeauthoryear{{Urban}, {Werner}, {Allen}, {Simionescu}  \&
  {Mantz}}{{Urban} et~al.}{2017}]{Urban:2017}
{Urban} O.,  {Werner} N.,  {Allen} S.~W.,  {Simionescu} A.,   {Mantz} A.,
  2017, \mn@doi [\mnras] {10.1093/mnras/stx1542}, \href
  {http://adsabs.harvard.edu/abs/2017MNRAS.470.4583U} {470, 4583}

\bibitem[\protect\citeauthoryear{{Vazza}, {Eckert}, {Simionescu}, {Br{\"u}ggen}
   \& {Ettori}}{{Vazza} et~al.}{2013}]{Vazza:2013aa}
{Vazza} F.,  {Eckert} D.,  {Simionescu} A.,  {Br{\"u}ggen} M.,   {Ettori} S.,
  2013, \mnras, 429, 799

\bibitem[\protect\citeauthoryear{{Wadsley}, {Stadel}  \& {Quinn}}{{Wadsley}
  et~al.}{2004}]{Wadsley:2004}
{Wadsley} J.~W.,  {Stadel} J.,   {Quinn} T.,  2004, \mn@doi [\na]
  {10.1016/j.newast.2003.08.004}, \href
  {http://adsabs.harvard.edu/abs/2004NewA....9..137W} {9, 137}

\bibitem[\protect\citeauthoryear{{Wadsley}, {Veeravalli}  \&
  {Couchman}}{{Wadsley} et~al.}{2008}]{Wadsley:2008}
{Wadsley} J.~W.,  {Veeravalli} G.,   {Couchman} H.~M.~P.,  2008, \mn@doi
  [\mnras] {10.1111/j.1365-2966.2008.13260.x}, \href
  {http://adsabs.harvard.edu/abs/2008MNRAS.387..427W} {387, 427}

\bibitem[\protect\citeauthoryear{{Wadsley}, {Keller}  \& {Quinn}}{{Wadsley}
  et~al.}{2017}]{Wadsley:2017}
{Wadsley} J.~W.,  {Keller} B.~W.,   {Quinn} T.~R.,  2017, \mn@doi [\mnras]
  {10.1093/mnras/stx1643}, \href
  {http://adsabs.harvard.edu/abs/2017MNRAS.471.2357W} {471, 2357}

\bibitem[\protect\citeauthoryear{{Walker} et~al.,}{{Walker}
  et~al.}{2019}]{Walker:2019}
{Walker} S.,  et~al., 2019, \mn@doi [\ssr] {10.1007/s11214-018-0572-8}, \href
  {http://adsabs.harvard.edu/abs/2019SSRv..215....7W} {215, 7}

\bibitem[\protect\citeauthoryear{Wang \& Walker}{Wang \&
  Walker}{2014}]{Wang:2014lr}
Wang Q.~D.,  Walker S.,  2014, MNRAS, 439, 1796

\bibitem[\protect\citeauthoryear{{Wang}, {Owen}  \& {Ledlow}}{{Wang}
  et~al.}{2004}]{Wang:2004}
{Wang} Q.~D.,  {Owen} F.,   {Ledlow} M.,  2004, \mn@doi [\apj]
  {10.1086/422332}, \href {http://adsabs.harvard.edu/abs/2004ApJ...611..821W}
  {611, 821}

\bibitem[\protect\citeauthoryear{{Werk}, {Prochaska}, {Thom}, {Tumlinson},
  {Tripp}, {O'Meara}  \& {Peeples}}{{Werk} et~al.}{2013}]{Werk:2013}
{Werk} J.~K.,  {Prochaska} J.~X.,  {Thom} C.,  {Tumlinson} J.,  {Tripp} T.~M.,
  {O'Meara} J.~M.,   {Peeples} M.~S.,  2013, \mn@doi [\apjs]
  {10.1088/0067-0049/204/2/17}, \href
  {http://adsabs.harvard.edu/abs/2013ApJS..204...17W} {204, 17}

\bibitem[\protect\citeauthoryear{{Werk} et~al.,}{{Werk}
  et~al.}{2014}]{Werk:2014}
{Werk} J.~K.,  et~al., 2014, \mn@doi [\apj] {10.1088/0004-637X/792/1/8}, \href
  {http://adsabs.harvard.edu/abs/2014ApJ...792....8W} {792, 8}

\bibitem[\protect\citeauthoryear{{Werner}, {Urban}, {Simionescu}  \&
  {Allen}}{{Werner} et~al.}{2013}]{Werner:2013}
{Werner} N.,  {Urban} O.,  {Simionescu} A.,   {Allen} S.~W.,  2013, \mn@doi
  [\nat] {10.1038/nature12646}, \href
  {http://adsabs.harvard.edu/abs/2013Natur.502..656W} {502, 656}

\bibitem[\protect\citeauthoryear{{White} \& {Rees}}{{White} \&
  {Rees}}{1978}]{White:1978}
{White} S.~D.~M.,  {Rees} M.~J.,  1978, \mn@doi [\mnras]
  {10.1093/mnras/183.3.341}, \href
  {http://adsabs.harvard.edu/abs/1978MNRAS.183..341W} {183, 341}

\bibitem[\protect\citeauthoryear{{Yoon} \& {Putman}}{{Yoon} \&
  {Putman}}{2013}]{Yoon:2013}
{Yoon} J.~H.,  {Putman} M.~E.,  2013, \mn@doi [\apjl]
  {10.1088/2041-8205/772/2/L29}, \href
  {http://adsabs.harvard.edu/abs/2013ApJ...772L..29Y} {772, L29}

\bibitem[\protect\citeauthoryear{Yoon \& Putman}{Yoon \&
  Putman}{2017}]{Yoon:2017aa}
Yoon J.~H.,  Putman M.~E.,  2017, The Astrophysical Journal, 839, 117

\bibitem[\protect\citeauthoryear{Yoon, Putman, Thom, Chen  \& Bryan}{Yoon
  et~al.}{2012}]{Yoon:2012yu}
Yoon J.~H.,  Putman M.~E.,  Thom C.,  Chen H.-W.,   Bryan G.~L.,  2012, ApJ,
  754, 84

\bibitem[\protect\citeauthoryear{{Zhuravleva}, {Churazov}, {Kravtsov}, {Lau},
  {Nagai}  \& {Sunyaev}}{{Zhuravleva} et~al.}{2013}]{Zhuravleva2013}
{Zhuravleva} I.,  {Churazov} E.,  {Kravtsov} A.,  {Lau} E.~T.,  {Nagai} D.,
  {Sunyaev} R.,  2013, \mn@doi [\mnras] {10.1093/mnras/sts275}, \href
  {http://adsabs.harvard.edu/abs/2013MNRAS.428.3274Z} {428, 3274}

\bibitem[\protect\citeauthoryear{{Zinger}, {Dekel}, {Birnboim}, {Kravtsov}  \&
  {Nagai}}{{Zinger} et~al.}{2016}]{Zinger2016}
{Zinger} E.,  {Dekel} A.,  {Birnboim} Y.,  {Kravtsov} A.,   {Nagai} D.,  2016,
  \mn@doi [\mnras] {10.1093/mnras/stw1283}, \href
  {http://adsabs.harvard.edu/abs/2016MNRAS.461..412Z} {461, 412}

\bibitem[\protect\citeauthoryear{{Zinger}, {Dekel}, {Kravtsov}  \&
  {Nagai}}{{Zinger} et~al.}{2018}]{Zinger:2018aa}
{Zinger} E.,  {Dekel} A.,  {Kravtsov} A.~V.,   {Nagai} D.,  2018, MNRAS, 475,
  3654

\bibitem[\protect\citeauthoryear{{van de Voort}, {Schaye}, {Booth}  \& {Dalla
  Vecchia}}{{van de Voort} et~al.}{2011}]{vandeVoort:2011}
{van de Voort} F.,  {Schaye} J.,  {Booth} C.~M.,   {Dalla Vecchia} C.,  2011,
  \mn@doi [\mnras] {10.1111/j.1365-2966.2011.18896.x}, \href
  {https://ui.adsabs.harvard.edu/\#abs/2011MNRAS.415.2782V} {415, 2782}

\makeatother
\end{thebibliography}


\end{document}